\documentclass[12pt]{article}
\usepackage{epsf}
\usepackage{amsmath}
\usepackage{graphics}

\usepackage{graphicx}
\input{colordvi.tex}

\renewcommand{\thefootnote}{\#\arabic{footnote}}

\newcommand{\lesssim}{ \mathop{}_{\textstyle \sim}^{\textstyle <} }

\begin{document}

\setcounter{footnote}{0}
\begin{titlepage}

\begin{center}

\hfill astro-ph/yymmnnn\\
\hfill March 2007\\

\vskip .5in

{\Large \bf
Constraints on Generalized Dark Energy from Recent Observations
}

\vskip .45in

{\large
Kiyotomo Ichiki$^1$ and Tomo Takahashi$^2$
}

\vskip .45in

{\em
$^1$Research Center for the Early Universe, the University of Tokyo, 
Tokyo 113-0033, Japan\\
$^2$
Department of Physics, Saga University, Saga 840-8502, Japan \\
}

\end{center}

\vskip .4in

\begin{abstract}

 Effects of a generalized dark energy fluid is investigated on cosmic
 density fluctuations such as cosmic microwave background.  As a
 general dark energy fluid, we take into consideration the possibility
 of the anisotropic stress for dark energy, which has not been discussed
 much in the literature.  We comprehensively study its effects on the
 evolution of density fluctuations along with that of non-adiabatic
 pressure fluctuation of dark energy, then give constraints on such a
 generalized dark energy from current observations. We show that,
 though we cannot find any stringent limits on the anisotropic stress
 or the non-adiabatic pressure fluctuation themselves, the constrains
 on the equation of state of dark energy can be affected in some cases
 by the nature of dark energy fluctuation characterized by these properties.
 This may have important implications to the strategy to study the nature of
 dark energy.

\end{abstract}
\end{titlepage}

\renewcommand{\thepage}{\arabic{page}}
\setcounter{page}{1}
\renewcommand{\thefootnote}{\#\arabic{footnote}}

\section{Introduction}

Almost all cosmological observations today suggest that the expansion of
the present universe is accelerating. To account for the present cosmic
acceleration, one usually assumes an enigmatic component called dark
energy.  Although many authors have  investigated the dark energy
to date, we still do not know its nature yet.  In most researches of
dark energy, one often parameterizes dark energy with its equation of
state $w_X = p_X / \rho_X$ where $p_X$ and $\rho_X$ are pressure and
energy density of dark energy.  Current cosmological observations such
as cosmic microwave background (CMB), type Ia supernovae (SNeIa),
large scale structure (LSS) and so on can give a constraint on $w_X$.
When one puts constraint on the equation of state, it can be assumed to be
constant in time for simplicity.  However many models of dark energy
proposed so far has a time-dependent equation of state. Thus most of
recent study of dark energy accommodate its time dependence in some
way.  Although its time dependence can be complicated in general, it
is usually assumed a simple form such as $w_X = w_0 + w_1 (1 -a)$
\cite{Chevallier:2000qy,Linder:2002et} when one tries to limit
$w_X$ accommodating its time dependence.  Some authors have also
discussed the constraint on dark energy using the time evolution of
the energy density of dark energy itself or Hubble parameter by
binning to divide them into several epochs \cite{Huterer:2004ch}.

In addition to the time dependence of the equation of state, there is
another property of dark energy of which we have to take care.  In
fact, in various models of dark energy, dark energy can fluctuate and
then its fluctuation can affect the cosmic density perturbations such
as CMB and LSS.  When one specifies a model for dark energy, one can
investigate its effects of fluctuation using the perturbation equations
in the model.  However, because there are many models proposed to date,
one 
may prefer to describe it in a phenomenological way.  To describe the
nature of fluctuation of dark energy, the (effective) speed of sound
$c_s^2$ has been used in many works so far.  Some authors have
investigated the constraints on the equation of state varying the
speed of sound \cite{Weller:2003hw} and on the speed of sound itself
\cite{Bean:2003fb,Hannestad:2005ak,Xia:2007km}.  Since the fluctuation of dark
energy mostly come into play after when dark energy becomes the
dominant component of the universe, the effects of it appear on large
scales.  Hence, when one tries to limit the speed of sound from
observations of CMB, one does not obtain a severe constraint on it
because the cosmic variance error is large there.  However, some
studies have shown that the constrains on the equation of state can
vary with the assumption for the perturbation property of dark energy.
For example, when one assumes that there is no fluctuation of dark
energy, the constraint for the constant equation of state is $ w_X =
-0.941^{+0.087}_{-0.101}$ \cite{Tegmark:2006az}.  However, when one
takes into account the perturbation of dark energy, the constraint
becomes $w_X = -1.00^{+0.17}_{-0.19}$ where $c_s^2=1$ is assumed
\cite{Tegmark:2006az}.  Thus, in this sense, it is important to
investigate the properties of dark energy fluctuation from the
viewpoint not only from to reveal its fluctuation nature but also from
constraining the equation of state for dark energy.

In fact, to parameterize the perturbation property of dark energy, the
use of the speed of sound is not enough.  If one considers the
possibilities of imperfect fluid models for dark energy, one also needs 
to specify its anisotropic stress in some way.  In fact, there are
some models proposed which have such anisotropic stress
(for example, \cite{Amendola:2005cr}). 
There are also a few works which accommodate such possibilities from a
phenomenological point of view
\cite{Hu:1998kj,Capozziello:2005pa,Koivisto:2005mm}.  In 
particular, the authors of Ref.~\cite{Koivisto:2005mm} have discussed
the implications of the anisotropic stress to observations such as
CMB.  Although they investigated its effects in some detail,
cosmological constrains in such models was not given.  Since we 
have now precise measurements of cosmology, it is worth to investigate 
constraints for dark energy in such a general setting.

The purpose of the present paper is to discuss a generalized dark
energy including the possibilities of anisotropic stress and give the
constraints on cosmological parameters assuming such a generalized
dark energy.

The structure of this paper is as follows. In the next section, we
briefly review the formalism for the investigation of a generalized
dark energy.  The perturbation equations are also given there.  Then,
in section 3, we discuss the effects of dark energy fluctuation on
CMB, with particular attentions to the anisotropic stress of dark
energy.  In section 4, we present the constraints on dark energy from
current cosmological observations. The final section is devoted to
conclusions and discussions.

\section{Formalism}

In this section, we summarize the formalism to investigate a
generalized dark energy. Some detailed descriptions of this issue can
be found in Refs.~\cite{Hu:1998kj,Koivisto:2005mm}.

The background evolution of dark energy can be parameterized by its
equation of state $w_X = p_X/ \rho_X$. The evolution of the energy
density of dark energy is given by solving
\begin{equation}
\rho_X' = - 3 \mathcal{H} (1 + w_X) \rho_X~,
\end{equation}
where the prime denotes the derivative with respect to the conformal
time  and $\mathcal{H}= a'/a$ is the conformal Hubble parameter
with $a$ being the scale factor.  As far as the background evolution
is concerned, this equation is enough to specify the effect of dark energy 
on the evolution of the universe.

However, as mentioned in the introduction, dark energy component
itself can fluctuate and affect cosmic density fluctuations.  For
the fluctuation of dark energy fluid, we have to specify more in addition
to the equation of state. Here we briefly summarize what we need to
consider a general dark energy. In the rest of the paper, we follow
the notation of Ref.~\cite{Ma:1995ey} and work in the synchronous
gauge whose metric perturbations are denoted as $h$ and $\eta$
unless otherwise stated.  We also sometimes make use of the
gauge-invariant gravitational potentials, which appear as metric 
perturbations in the conformal Newtonian gauge,  to discuss the effects of
fluctuation of dark energy.

To specify the nature of dark energy perturbation, we need to give its
pressure perturbation and anisotropic stress.  For the pressure
perturbation, it may be useful to separate it to adiabatic and
non-adiabatic parts.  When the dark energy fluid is adiabatic, the
evolutions of pressure perturbation can be specified with the
adiabatic sound speed $c_a^2$ which is written as
\begin{equation}
c_a^2 \equiv \frac{p_X'}{\rho_X' } 
= 
w_X - \frac{w_X'}{3 \mathcal{H} ( 1 + w_X )}.
\end{equation}
Then pressure perturbation is given as $\delta p = c_a^2 \delta \rho$.
If $w_X<0$ and the time evolution of $w_X$ is small compared with the
Hubble expansion rate, $c_a^2$ can be negative and the adiabatic pressure can
not support density fluctuations.
However, in general, non-adiabatic pressure fluctuation may arise.
Such a degrees of freedom can be parameterized with the so-called
(effective) speed of sound which is defined as
\begin{equation}
c_s^2 \equiv \frac{ \delta p_X}{\delta \rho_X} \Bigl|_{\rm rest}.
\end{equation}
This quantity is usually specified in the rest frame of dark energy
and we follow the convention here.  A famous example of dark
energy models with non-adiabatic pressure fluctuation is a scalar
field, called quintessence, 
which has the speed of sound $c_s^2=1$.  Other models such as k-essence, 
a scalar field model which can have a non-canonical kinetic term,  can have
different values for the speed of sound.  In the adiabatic case,
$c_s^2$ and $c_a^2$ coincide.

In fact, to consider a general fluid model for dark energy, the speed
of sound is not enough to specify its nature.  We still have to
determine an anisotropic stress for dark energy.  Including the
anisotropic stress, perturbation equations for density and velocity
fluctuations are
\begin{eqnarray}
\delta_X' 
&=& 
-( 1 + w_X) 
\left[ k^2 + 9 \mathcal{H}^2 ( c_s^2 - c_a^2 )\right] \frac{\theta_X}{k^2} 
- 3 \mathcal{H} (c_s^2 - w_X ) \delta_X
-( 1 + w_X) \frac{h'}{2} 
\label{eq:delta_X}\\
\theta_X' 
&=& 
-\mathcal{H} ( 1 - 3c_s^2 ) \theta_X 
+ \frac{c_s^2 k^2}{1 +w_X} \delta_X
-k^2 \sigma_X
\label{eq:theta_X}
\end{eqnarray}
where $\delta_X, \theta_X $ and $\sigma_X$ represent density, velocity
and anisotropic stress perturbations\footnote{
The relation between $\sigma_X$ here and $\pi_g$ in Ref.~\cite{Hu:1998kj} is 
\[
\sigma_X = \frac{2}{3} \frac{w_X}{1 + w_X} \pi_g
\]
} for a general fluid, respectively.  The anisotropic stress can be
presented as the viscosity and damps velocity perturbation in
sheer-free frame.  Thus, in addition to the sound speed of dark energy
$c_s^2$, there is another freedom to specify the anisotropic stress
$\sigma_X$ of dark energy.  In fact, it may be possible to choose
$\sigma_X$ demanding that it is consistent with viscous damping of
velocity perturbation in sheer-free frame. However, in this paper we
follow a phenomenological approach of
Ref.~\cite{Hu:1998kj,Koivisto:2005mm}.  Considering the transformation
property between frames, $\sigma_X$ should be gauge invariant thus its evolution
can be given by the gauge-invariant combinations of some metric and
velocity shear. Taking into account the dissipation, the anisotropic
stress can be obtained by solving \cite{Hu:1998kj,Koivisto:2005mm}
\begin{equation}
\sigma_X' + 3 \mathcal{H} \frac{c_a^2}{w_X} \sigma_X
=
\frac{8}{3} \alpha
\left( \theta_X + \frac{h'}{2} + 3 \eta' \right)~.
\label{eq:sigma_X}
\end{equation}
Here $\alpha$ is the viscosity parameter which specifies the
nature of an anisotropic stress for dark energy.  In fact, in some
literatures, another parameter $c_{\rm vis}^2$ is called the viscosity
parameter, which is related to $\alpha$ as
\begin{equation}
\label{eq:alpha}
\alpha =  \frac{c_{\rm vis}^2}{1 + w_X}.
\end{equation}
However, in this paper, we use $\alpha$ as the viscosity parameter.
It should be mentioned that $\alpha$ must be kept positive to preserve
the physical effects where the viscosity damps the perturbations.
Thus, in the following we consider the case with $\alpha > 0$.

We modified CAMB code \cite{Lewis:1999bs} to include a general dark
energy fluid to calculate CMB and matter power spectra.  We also checked our
results by modifying CMBFAST \cite{Seljak:1996is} and found that they both
give the same results with good accuracy.

\section{Effects of dark energy fluctuation}
\label{sec:effect}

In this section, we discuss the effects of dark energy fluctuation,
paying particular attentions to the anisotropic stress of dark energy.
Since the dark energy can be a dominant component of the universe only
at late times, the effects of dark energy fluctuation is significant
on large scales.\footnote{
  The background evolution of dark energy can affect the expansion of
  the universe, thus the peak positions of CMB power spectrum can
  depend on the equation of state of dark energy. However, as far as
  the fluctuation of dark energy concerned, its effects can be large
  only on large scales unless the equation of state dark energy is close to 
  $0$ at early times.
  }
In particular, as we will show in the following, the effects of the
anisotropic stress are significant at low multipoles of CMB power
spectrum, through the late time integrated Sachs-Wolfe
(ISW) effect.  Since the ISW effect is driven by the change of the
gravitational potential, we start our discussion with investigating
the gravitational potential by changing the viscosity parameter 
$\alpha$.  In fact, as it has been discussed in the literature
\cite{Pietrobon:2006gh,Hu:2004yd}, the non-adiabatic pressure
fluctuation, which is usually 
characterized by the effective speed of sound $c_s^2$, can also affect
the late time ISW effect. Thus we also discuss the effects of the
non-adiabatic pressure fluctuation along with that of the anisotropic
stress. 

\begin{figure}[ht]
\begin{minipage}[m]{0.5\linewidth}
\rotatebox{0}{\includegraphics[width=0.9\textwidth]{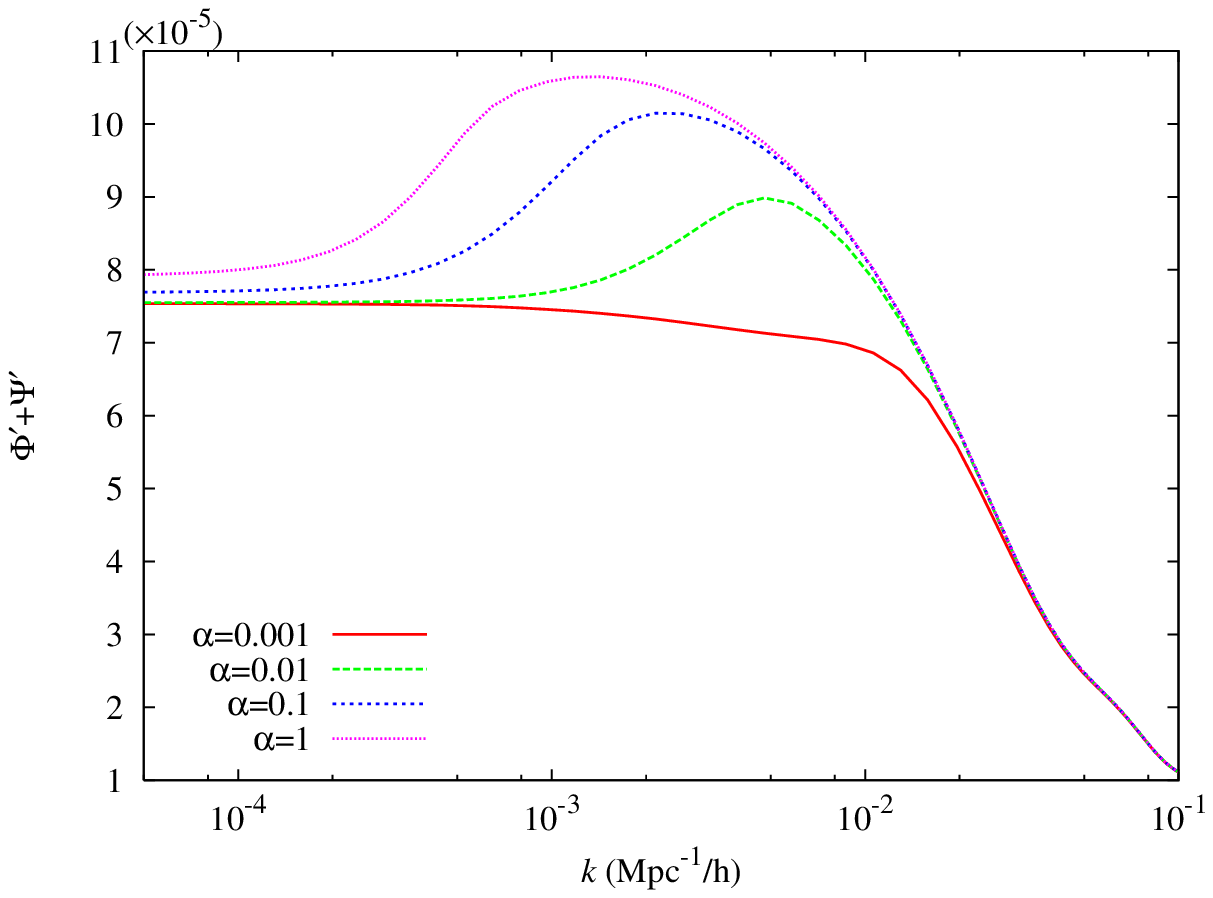}}
\end{minipage}
\begin{minipage}[m]{0.5\linewidth}
\rotatebox{0}{\includegraphics[width=0.9\textwidth]
{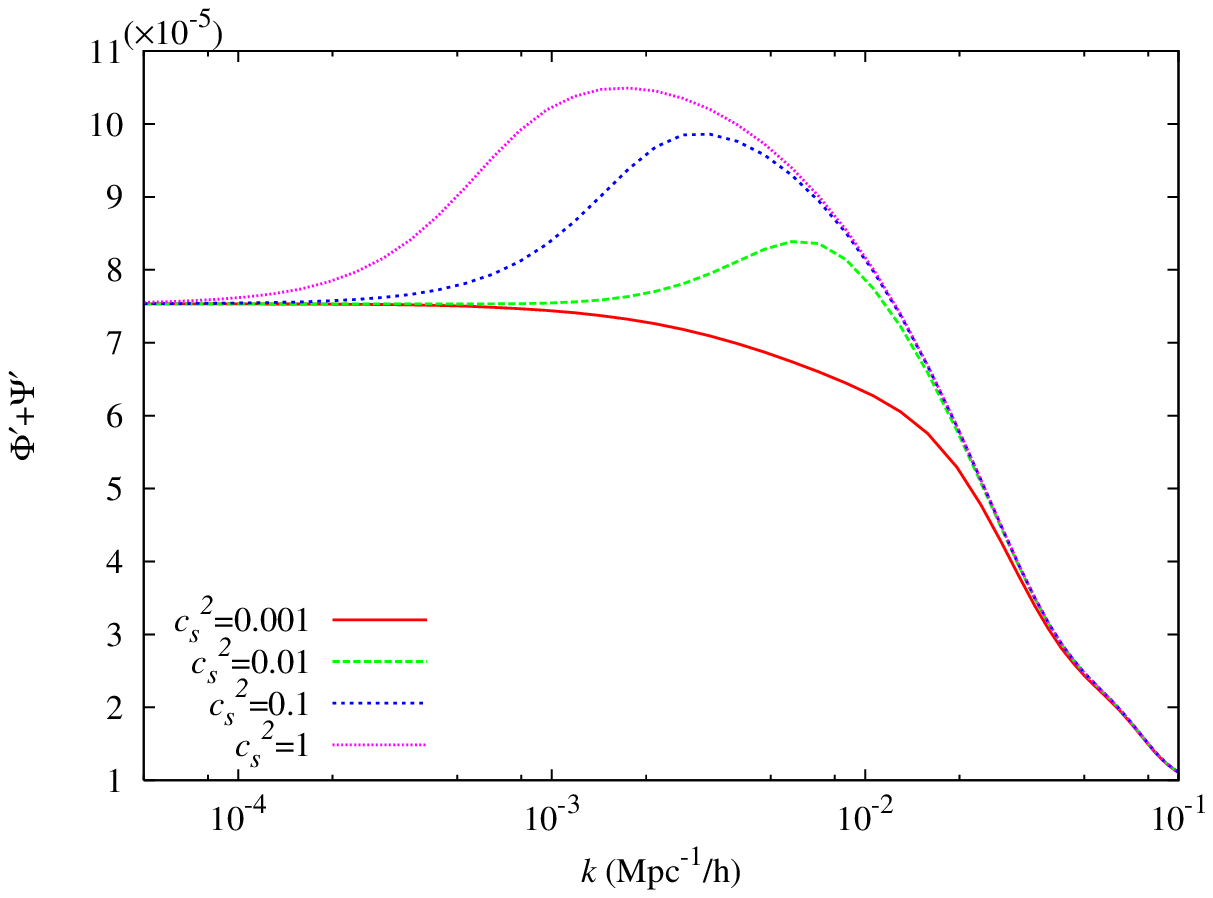}}
\end{minipage}
\caption{The value of $\Phi' + \Psi'$ today is shown as a function of
  $k$ for several values of $\alpha$ (left) and $c_s^2$ (right).  The
  equation of state is fixed as $w_X=-0.8$ in this figure.  }
\label{fig:1}
\end{figure}

Since we are going to discuss the effects on the ISW effect, first we
show time derivative of the gravitational potential $\Phi' + \Psi'$ at
present time in Fig.~\ref{fig:1} as a function of wave number $k$ for
several values of $\alpha$ and $c_s^2$. Here and hereafter gravitational
potentials are normalized at super horizon scales such that $\Psi =
-10/(4R_\nu+15)$ with $R_\nu=\rho_\nu/(\rho_\gamma+\rho_\nu)$ being
neutrino fraction, and the conformal time coordinate is in units of Mpc following the
convention. 
 The equation of state for
dark energy is assumed as $w_X=-0.8$ in
this figure. When we vary $\alpha$, $c_s^2$ is
assumed to be $0$ and vice versa.  For illustrative purpose, other
cosmological parameters are fixed as $\Omega_mh^2=0.128$ (dark matter density), $\Omega_bh^2=
0.022$ (baryon density), $h=0.73$ (hubble parameter), $\tau=0.09$
(optical depth to the last scattering surface) and $n_s=0.96$ (spectral
index of primordial density fluctuation), which are mean values 
for $\Lambda$CDM model from the analysis using WMAP3 data alone 
\cite{Spergel:2006hy}.  In this
paper, we assume that the universe is flat and no running of scalar
spectral index and no tensor mode contribution.  Unless otherwise
stated, we use the above cosmological parameters except for those of
dark energy.  As seen from the figure, the change of the gravitational
potential increases as $\alpha$ and $c_s^2$ becomes larger.  Since
$\alpha$ represents the strength of the anisotropic stress, which
damps velocity perturbations, the increase of $\alpha$ drives the change
of the gravitational potential. Thus we obtain larger ISW effect as
$\alpha$ increases.  As for the effects of the non-adiabatic pressure
fluctuation, $c_s^2$ determines
the sound horizon $s = \int c_s dt/a$ under which dark energy can be
assumed as ``smooth" component.  When dark energy is considered to be
smooth, fluctuation of dark energy can act as stabilizing force for
the gravitational potential, which leads to the decay of the
gravitational potential. Above the sound horizon, dark energy can
cluster, thus only gravitational force can be relevant to the
evolution of the gravitational potential. The transition between
``smooth" regime to clustered regime occurs at larger scales as $c_s^2$
increases. Thus the larger values of $c_s^2$ cause the more late time ISW
effect, which can be observed in the figure.

\begin{figure}[ht]
\begin{minipage}[m]{0.5\linewidth}
\rotatebox{0}{\includegraphics[width=0.9\textwidth]{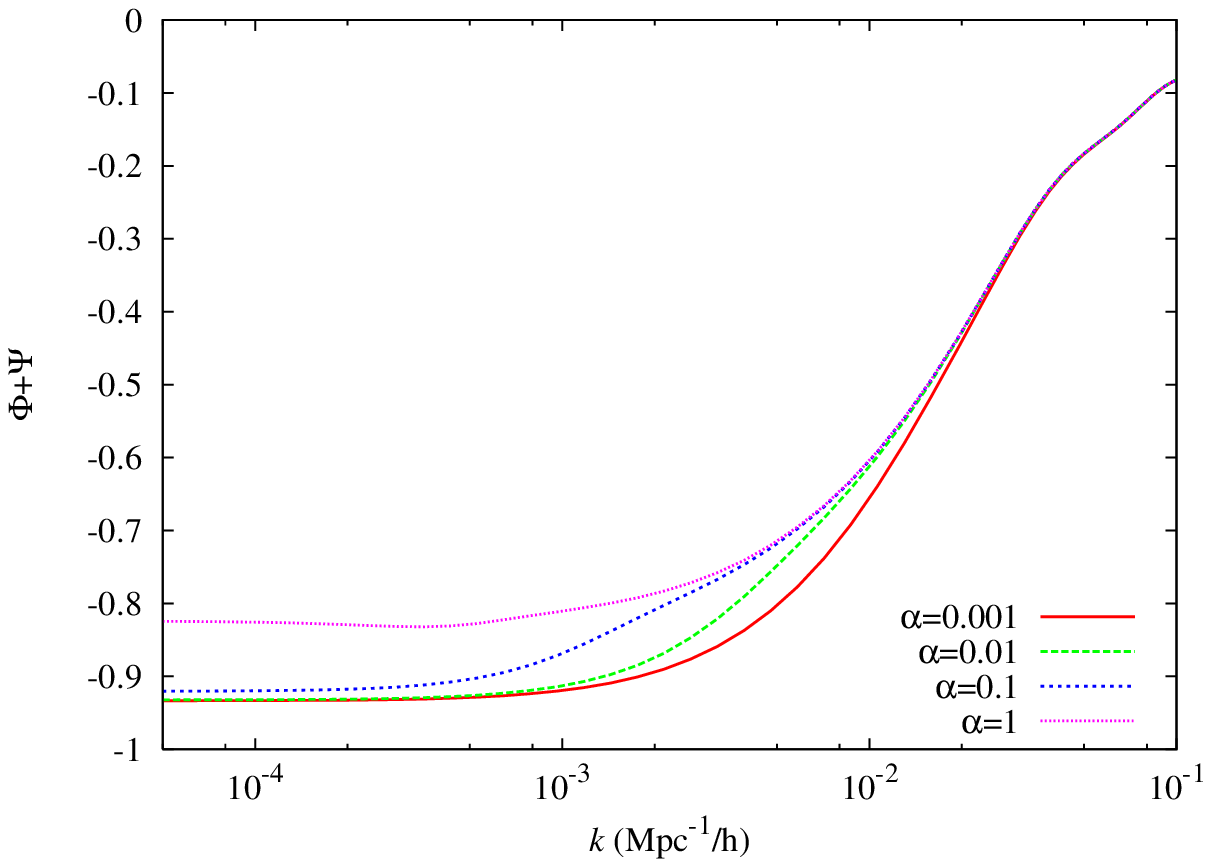}}
\end{minipage}
\begin{minipage}[m]{0.5\linewidth}
\rotatebox{0}{\includegraphics[width=0.9\textwidth]{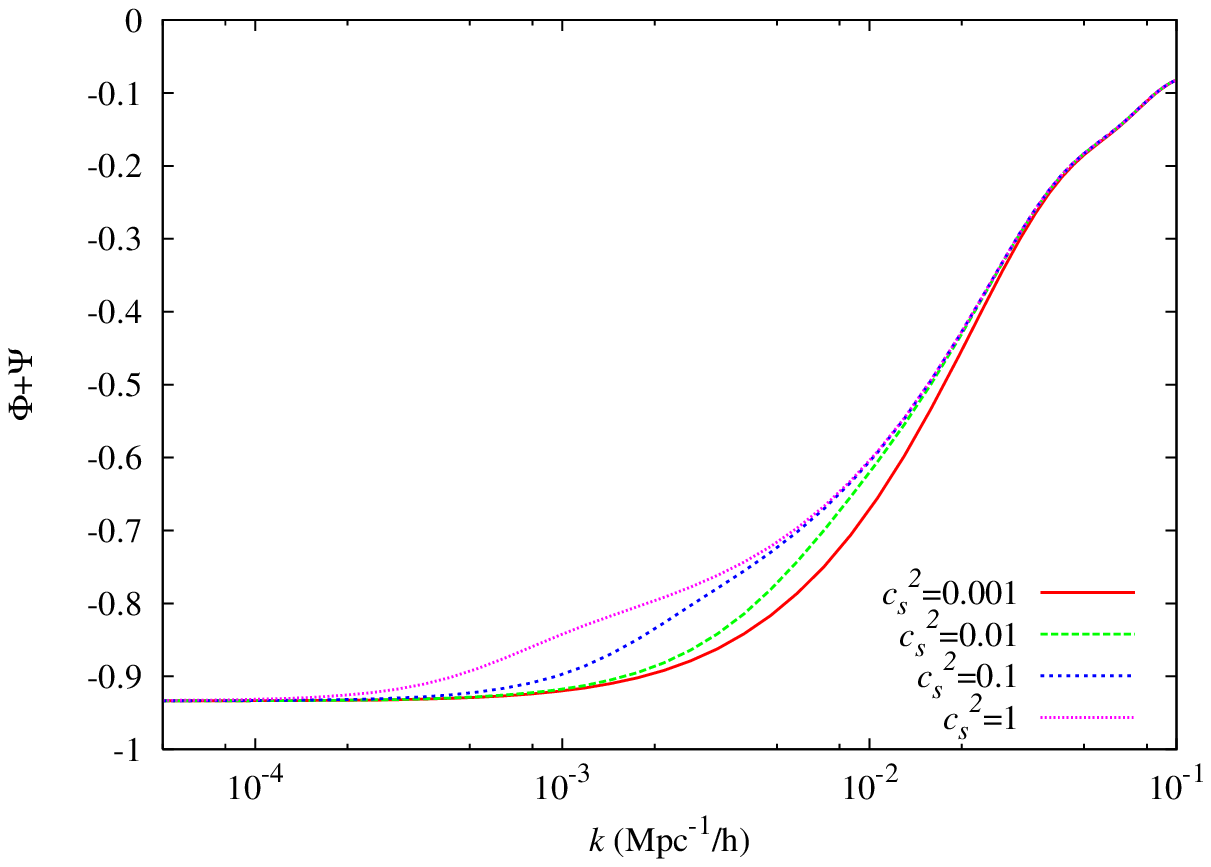}}
\end{minipage}
\caption{The value of $\Phi + \Psi$ today is shown as a function of
  $k$ for several values of $\alpha$ (left) and $c_s^2$ (right).  The
  equation of state is fixed as $w_X=-0.8$ in this figure. }
\label{fig:2}
\end{figure}

For the effects of the non-adiabatic pressure fluctuation, the behavior
of the gravitational potential differs depending on $c_s^2$ only around the 
transition regime but not at smooth or clustered limit. However, for the
anisotropic stress, the ISW effect depends on $\alpha$ on large
scales. To see this clearly, we show the present value of $\Phi +
\Psi$ as a function of $k$ in Fig.~\ref{fig:2}.  For the
case with $\alpha$ being fixed and varying $c_s^2$, it can be clearly
seen that the behaviors on large/small scales are the same and they
differ only at transition regime.  However, for the case with $c_s^2$
being fixed and varying $\alpha$, corresponding lines depend on the
value of $\alpha$ even on large scale limit. 
 Since the parameter $\alpha$ determines the
magnitude of the anisotropic stress $\sigma_X$ through
Eq.~\eqref{eq:sigma_X}, the change of the gravitational potential is
also affected by the magnitude of $\alpha$.  
On the other hand, $c_s^2$ only determines
the sound horizon which differentiates clustering property of dark
energy, thus the magnitude of $c_s^2$ itself is not directly relevant
to the amount of the change of the gravitational potential. This is
one of the differences between the effects of $c_s^2$ and $\alpha$ although
they affect fluctuation on large scale in a similar way.

\begin{figure}[ht]
\begin{minipage}[m]{0.5\linewidth}
\vspace*{0.4cm}
\rotatebox{0}{\includegraphics[width=0.95\textwidth]{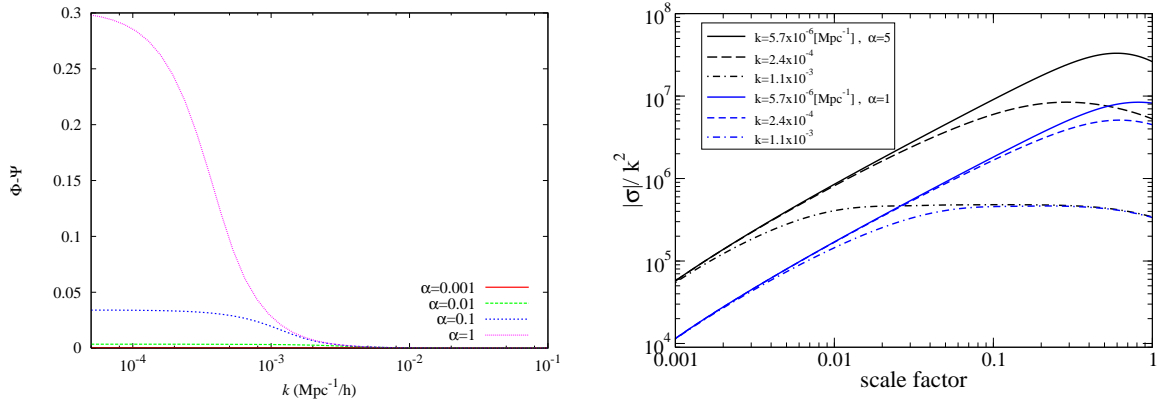}}
\end{minipage}
\begin{minipage}[m]{0.5\linewidth}
\vspace*{0.4cm}
\rotatebox{0}{\includegraphics[width=0.90\textwidth]{fig3-2.eps}}
\end{minipage}
 \caption{(Left) Scale dependence of $\Phi- \Psi$ at the present time are
   plotted assuming several values for $\alpha$. In this figure,
   $c_s^2$ are set to be $0$. (Right) Time evolutions of $\sigma_X$ for
 viscosity parameters $\alpha=5$ (black lines) and $\alpha=1$ (blue
 lines). Different lines correspond to the evolutions of different
 Fourier modes as indicated in the figure. The larger viscosity parameter
 makes the evolution of $\sigma$ after horizon crossing the slower
 during the matter dominated era. After the dark energy dominates
 the energy density of the universe, the perturbation starts to be erased.} 
 \label{fig:3}
\end{figure}

Furthermore, it is well-known that $\Phi$ and $\Psi$ coincide when
there is no anisotropic stress, in other words, $\Phi - \Psi$ is
sensitive to the anisotropic stress perturbation $\sigma$.  This can
be understood from the perturbed Einstein equation, as
\begin{equation}
k^2 \left( \Phi - \Psi \right) 
=
12 \pi G_N a^2 ( \rho + p) \sigma
\label{eq:Phi_Psi}
\end{equation}
where the quantity in the RHS is for the total matter.  In the left
panel of Fig.~\ref{fig:3}, the values of $\Phi - \Psi$ at the present
time are shown for several values of $\alpha$ as a function of wave
number.  As seen from the figure, $|\Phi -\Psi|$ becomes larger as
$\alpha$ increases on large scales, which can be easily understood
since larger values of $\alpha$ means large anisotropic stresses from
dark energy.  Another point which should be noticed is that the
effect of it becomes larger on large scales. For the generalized dark
energy with non-zero $\alpha$, the viscosity acts against the
gravitational collapse once the modes come across the horizon (right
panel of Fig. \ref{fig:3}).
Therefore the perturbations including the anisotropic stress
$\sigma$ have damped more at smaller scales, which makes the effect
significant only at large scales at present.

\begin{figure}[htb]
\begin{minipage}[m]{0.5\linewidth}
\rotatebox{0}{\includegraphics[width=0.9\textwidth]{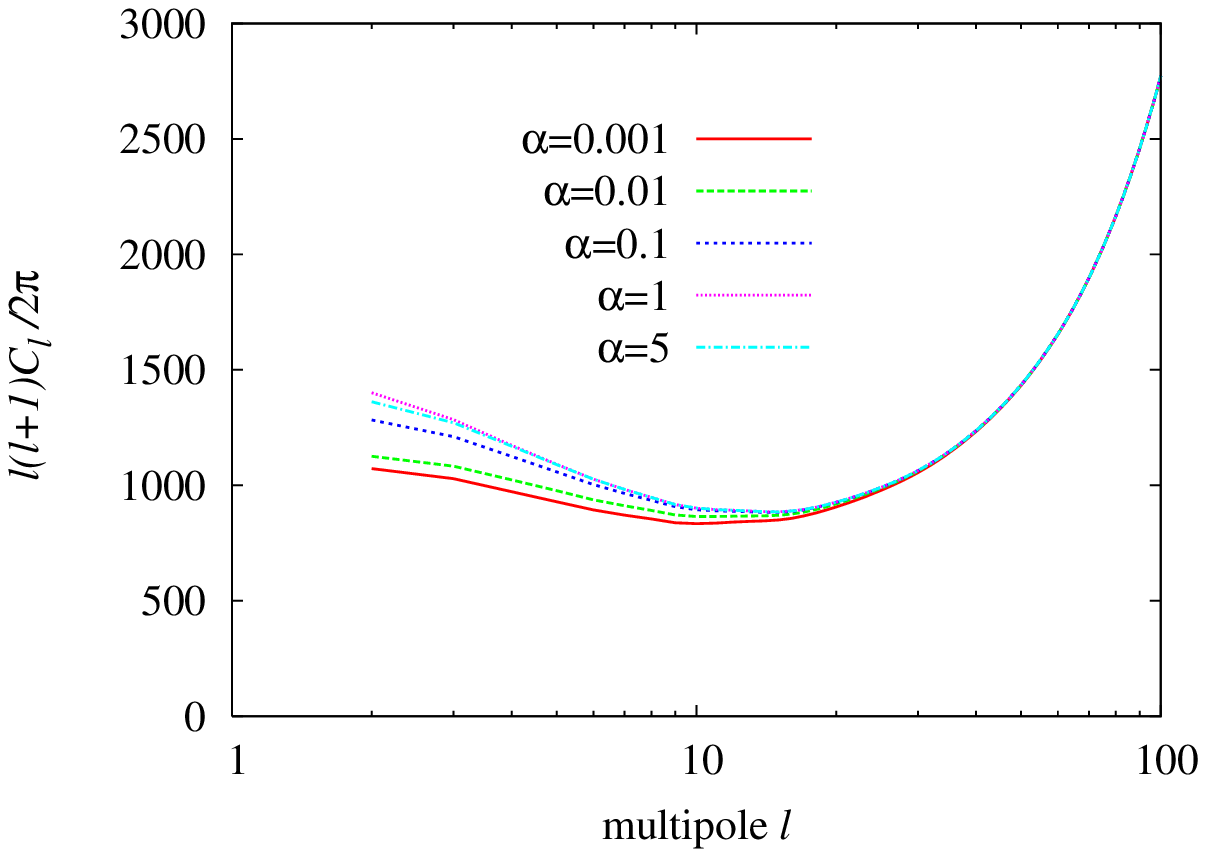}}
\end{minipage}
\begin{minipage}[m]{0.5\linewidth}
\rotatebox{0}{\includegraphics[width=0.9\textwidth]{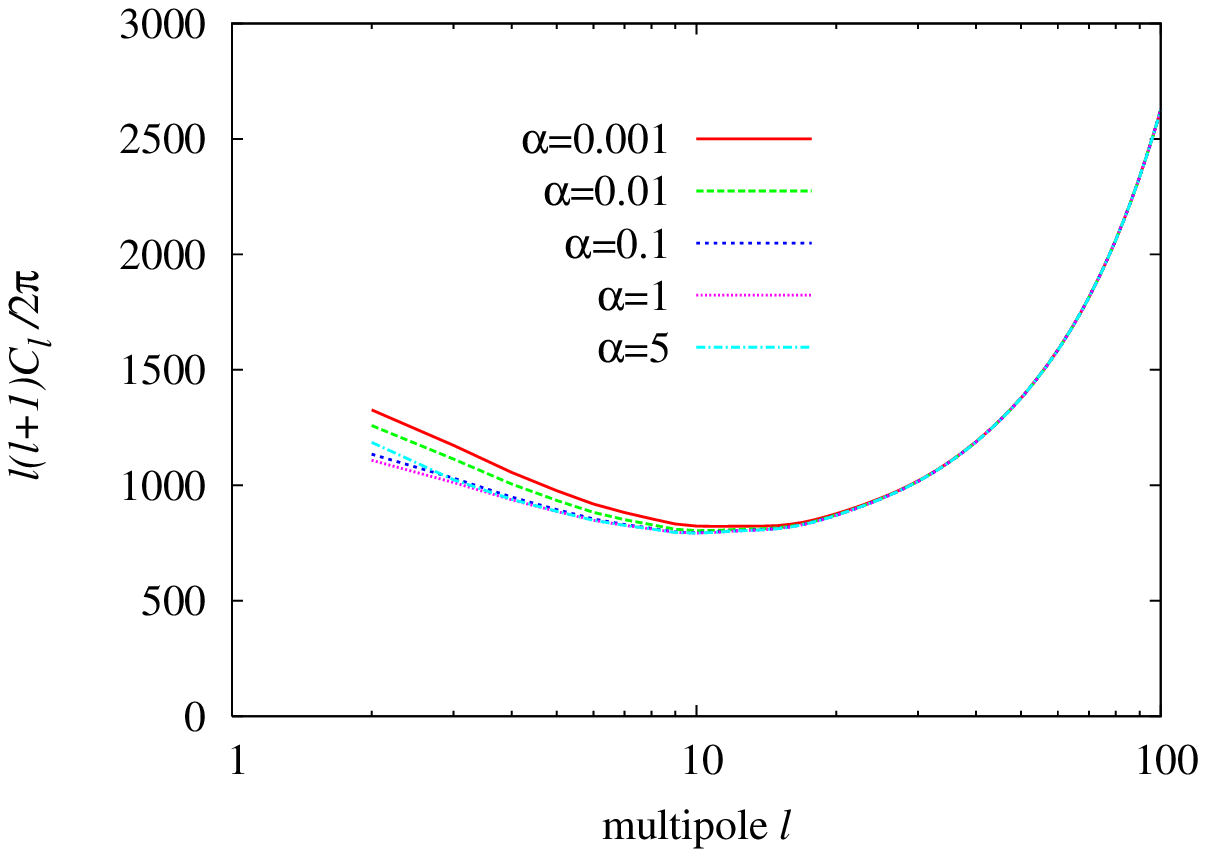}}
\end{minipage}
\begin{minipage}[m]{0.5\linewidth}
\rotatebox{0}{\includegraphics[width=0.9\textwidth]{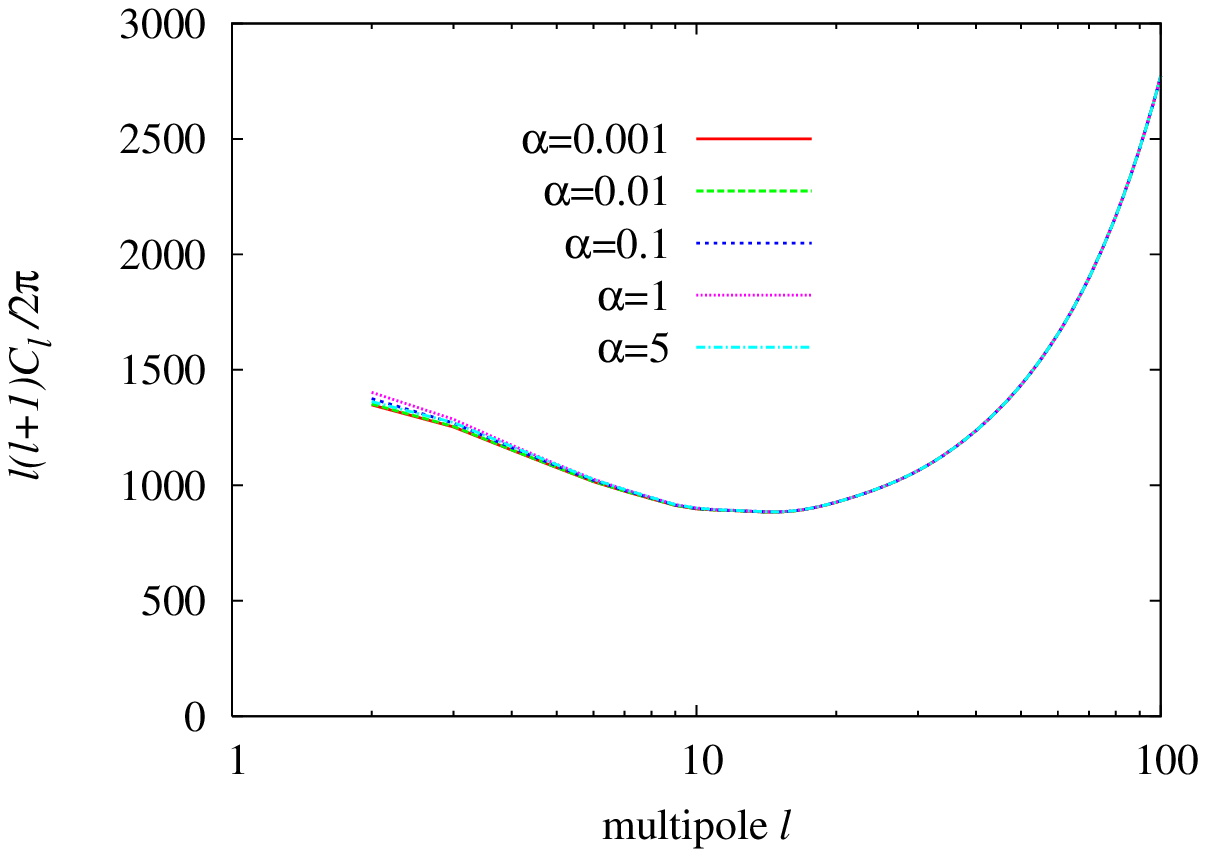}}
\end{minipage}
\begin{minipage}[m]{0.5\linewidth}
\rotatebox{0}{\includegraphics[width=0.9\textwidth]{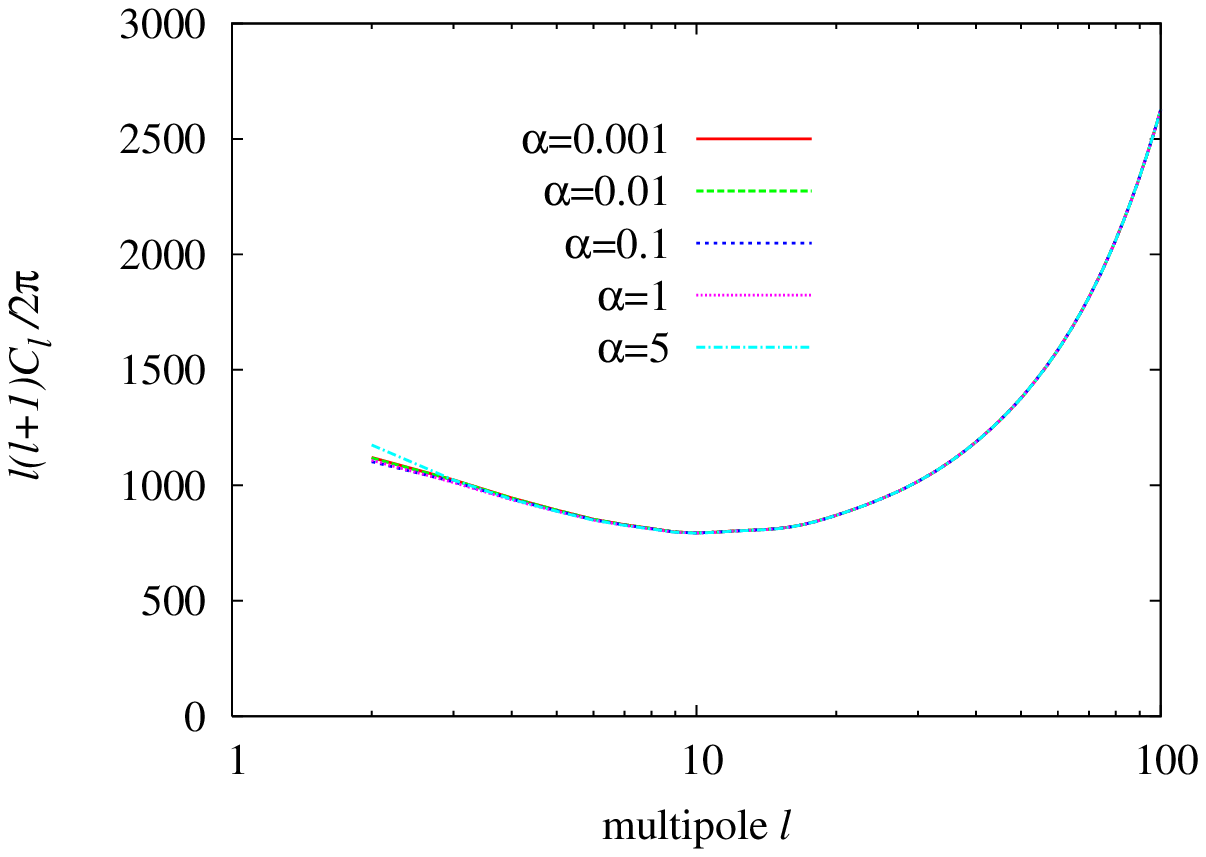}}
\end{minipage}
\caption{CMB TT power spectra for several values of $\alpha$ for the
  cases with $w_X=-0.8,  c_s^2= 0$ (top left), $w_X = -1.2, c_s^2=0$
  (top right), $w_X = -0.8, c_s^2=1$ (bottom right) and $w_X = -1.2,
  c_s^2=1$ (bottom right).}
\label{fig:4}
\end{figure}

\begin{figure}[htb]
\begin{minipage}[m]{0.5\linewidth}
\rotatebox{0}{\includegraphics[width=0.9\textwidth]{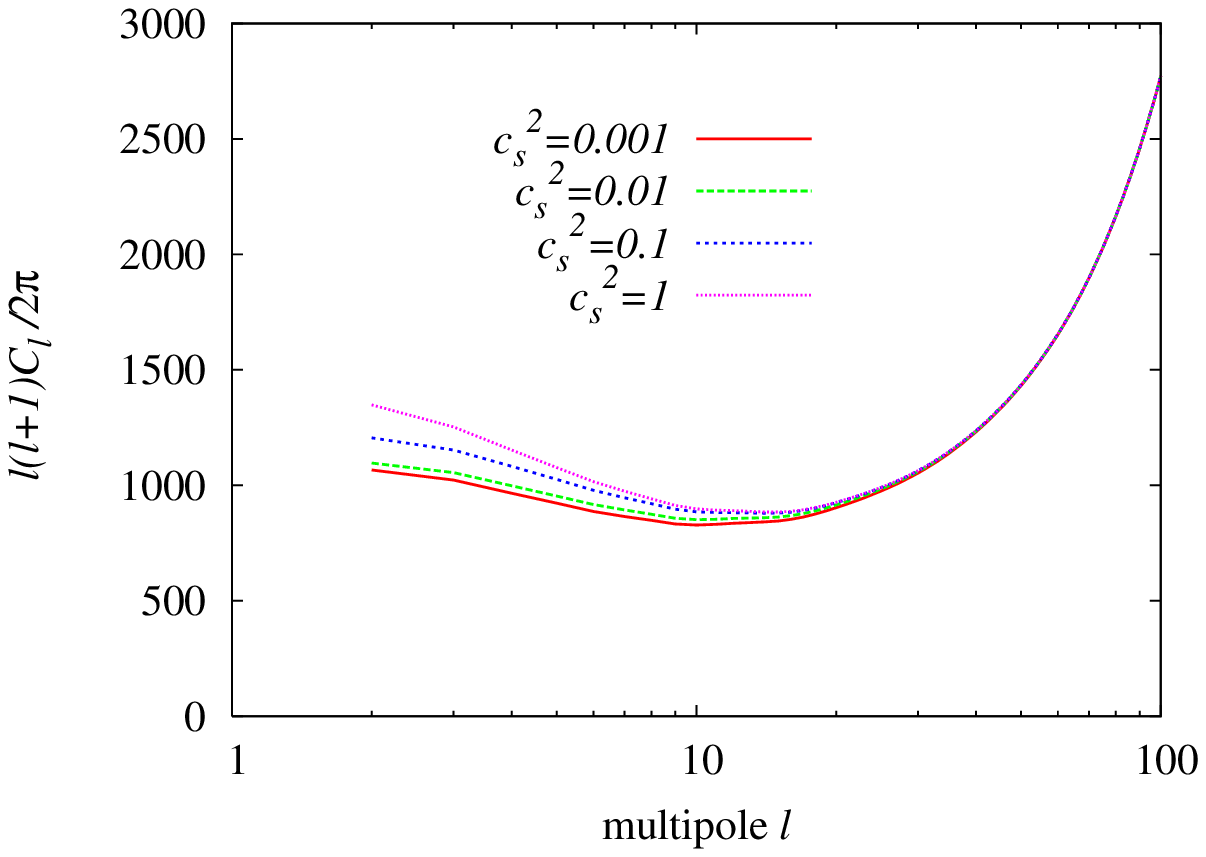}}
\end{minipage}
\begin{minipage}[m]{0.5\linewidth}
\rotatebox{0}{\includegraphics[width=0.9\textwidth]{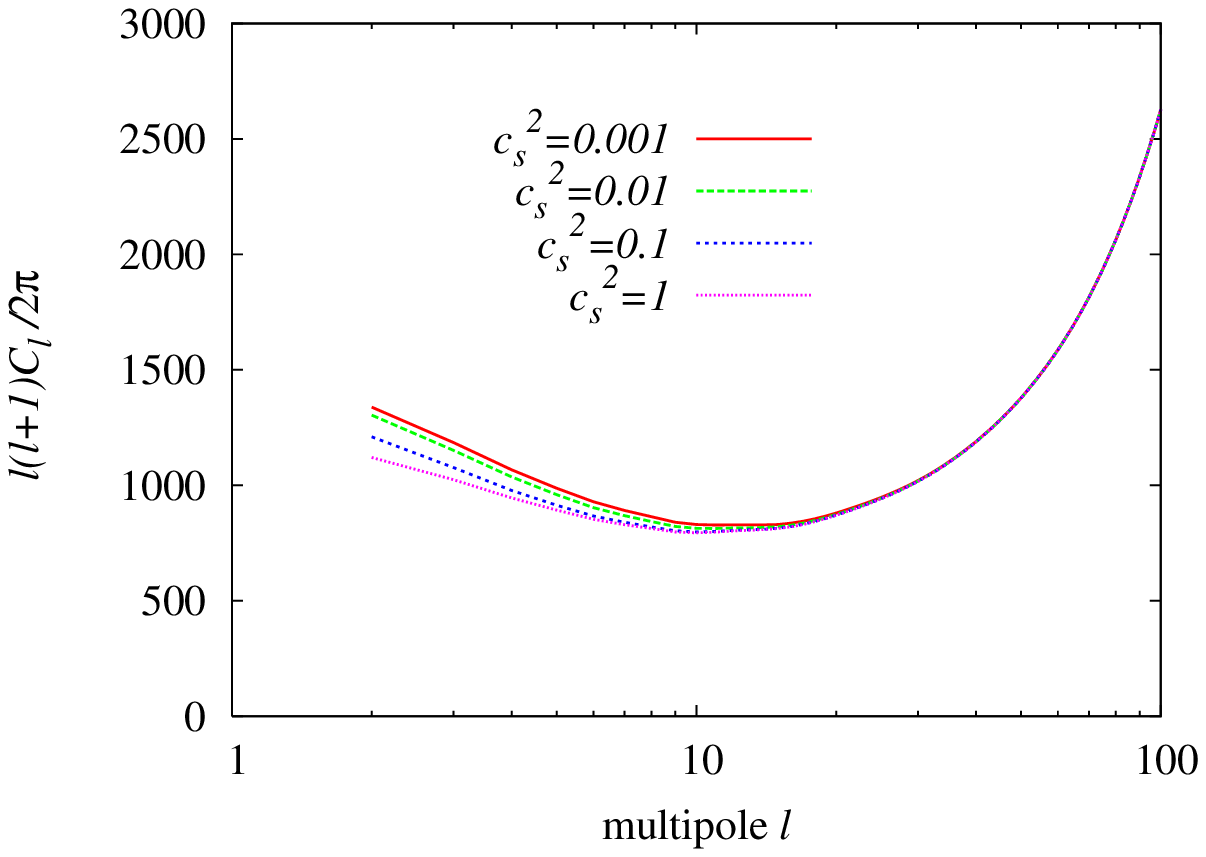}}
\end{minipage}
\begin{minipage}[m]{0.5\linewidth}
\rotatebox{0}{\includegraphics[width=0.9\textwidth]{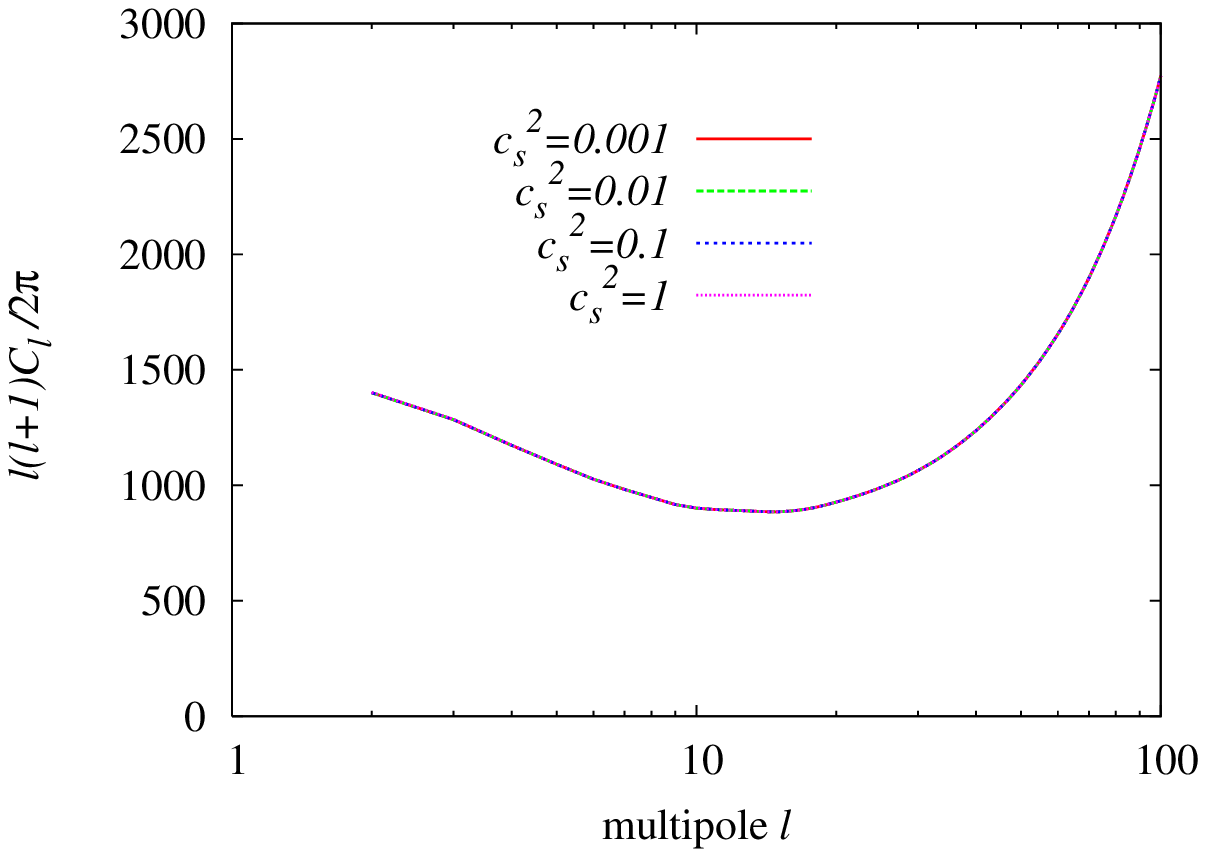}}
\end{minipage}
\begin{minipage}[m]{0.5\linewidth}
\rotatebox{0}{\includegraphics[width=0.9\textwidth]{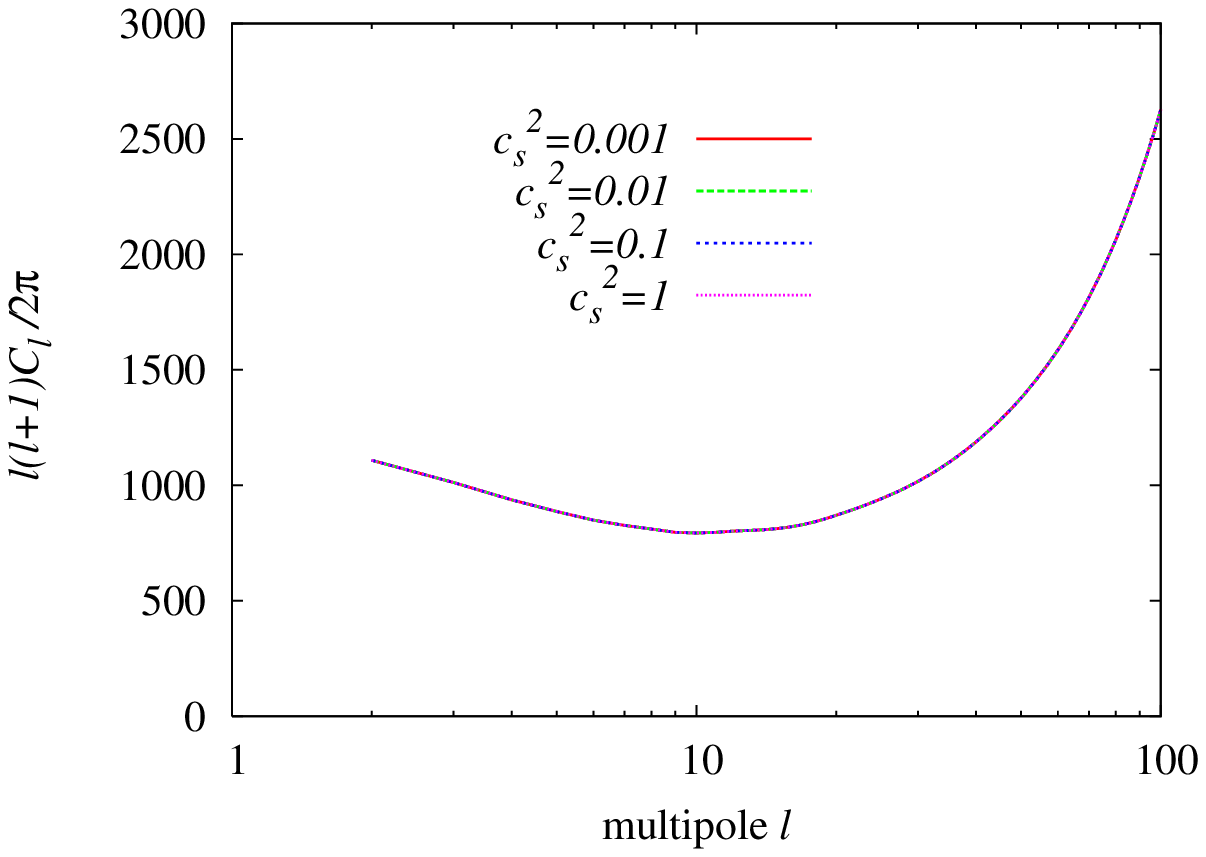}}
\end{minipage}
\caption{CMB TT power spectra for several values of $c_s^2$ for the
  cases with $w_X= -0.8, \alpha=0$ (top left), $w_X = -1.2, \alpha=0$
  (top right), $w_X = -0.8, \alpha=1$ (bottom right) and $w_X = -1.2,
  \alpha=1$ (bottom left).}
\label{fig:5}
\end{figure}

Now we are going to discuss the effects on CMB TT power spectra.  In
Figs.~\ref{fig:4} and \ref{fig:5}, the CMB TT power spectra
are depicted for several values of $\alpha$ and $c_s^2$.  The value
of the equation of state $w_X$ are found in the captions.  As noted
above, since the dark energy can be dominant component of the universe
only at late times, its fluctuation affects the CMB power spectra on
large scales.  Thus we only show low multipole region in the figures.
As seen from Fig.~\ref{fig:4}, $C_l$ increases as $\alpha$
increases for the case with $w_X=- 0.8$ and $c_s^2=0$.  This is
expected from the argument on the ISW effect above.  However, when the
equation of state is less than $-1$, the tendency is the opposite,
i.e., as the $\alpha$ increases, the large scale power decreases.  
This can be understood by looking at the source term in the 
perturbation equation.
The anisotropic stress can affect the evolution of the gravitational
potential through the change of density and velocity perturbations.
Such an effect is induced by the source term in Eq.~\eqref{eq:delta_X}
which depends on the prefactor $1+w_X$. Thus the effect should be
different depending on the sign of the factor $1+w_X$. Hence the tendency
becomes the opposite for  the cases with $w_X >-1$ and $<-1$.

When the value of the sound speed is fixed as $c_s^2=1$, the effects
of the anisotropic stress become insignificant for any value of
$w_X$. For the case with $c_s^2=1$, most scales relevant to low multipole 
region where dark energy
fluctuation can be important are in smooth regime.  As shown above, the anisotropic
stress can affect the gravitational potential, at least superficially,
similarly to smoothed dark energy.  Hence when dark energy can be assumed 
as a  smooth component,
the effect of the anisotropic stress becomes irrelevant since dark
energy is already smoothed.
 
\begin{figure}[htb]
\begin{center}
\rotatebox{0}{\includegraphics[width=0.8\textwidth]{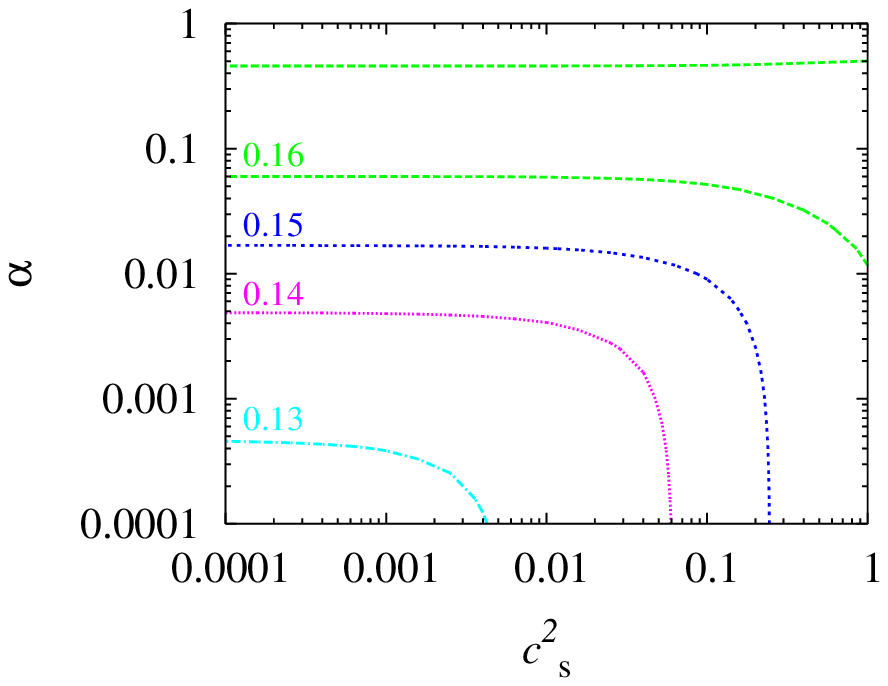}}
\end{center}
 \caption{Contours of the ratio $\tilde{C}_2/\tilde{C}_{\rm 1st}$ are
   plotted in the $c^2_s -\alpha$ plane for the case with $w_X
   =-0.8$. }
   \label{fig:6}
\end{figure}

For comparison, we also show the same plot for the effects of the
non-adiabatic pressure fluctuation.  In Fig.~\ref{fig:5}, CMB TT power
spectra 
are shown for several values of $c_s^2$ fixing $\alpha$ and other
parameters.  Since the effects of the anisotropic stress and non-adiabatic 
pressure fluctuation are similar in the ISW effect, 
the tendency should be  also the same as that of the effects of
$\alpha$.  As anticipated, in Fig.~\ref{fig:5}, we can see almost
the same behavior of low multipole region for the case with $c_s^2$
being varied, as for the case with $\alpha$ being varied.
Thus we can expect that there are some degeneracy between
the effects of $\alpha$ and $c_s^2$ on the fluctuation at low multipoles
of CMB.  As discussed above, they affect the ISW effect which
enhances/reduces the power at low multipoles depending on the strength
of these effects characterized by $\alpha$ and $c_s^2$.  Although the
origin of these effects are different in nature, they seem to affect
low multipoles of CMB in the same manner.  To see this quantitatively,
we show the contours of constant ratio of $\tilde{C}_l \equiv l(l+1) C_l$
at between $l=2$ and the first peak in Fig.~\ref{fig:6}.  In
the figure, the equation of state of dark energy is assumed as $w_X =
-0.8$. Other cosmological parameters are also fixed as the values already noted
above.  Notice that higher multipole region $l > \mathcal{O}(100)$ is
unchanged by  varying the values of $c_s^2$ and $\alpha$.  The
ratio represents how low multipoles are enhanced/reduced by the effects of the
anisotropic stress and non-adiabatic pressure fluctuation.  The figure
indicates that some combination of $c_s^2$ and $\alpha$ give the same
effect on 
large scales.  Thus when we study the constraint from observations, 
we can expect some degeneracy between $c_s^2$ and
$\alpha$.  We discuss this issue in the next section.


\begin{figure}[htb]
\begin{minipage}[m]{0.5\linewidth}
\rotatebox{0}{\includegraphics[width=0.9\textwidth]{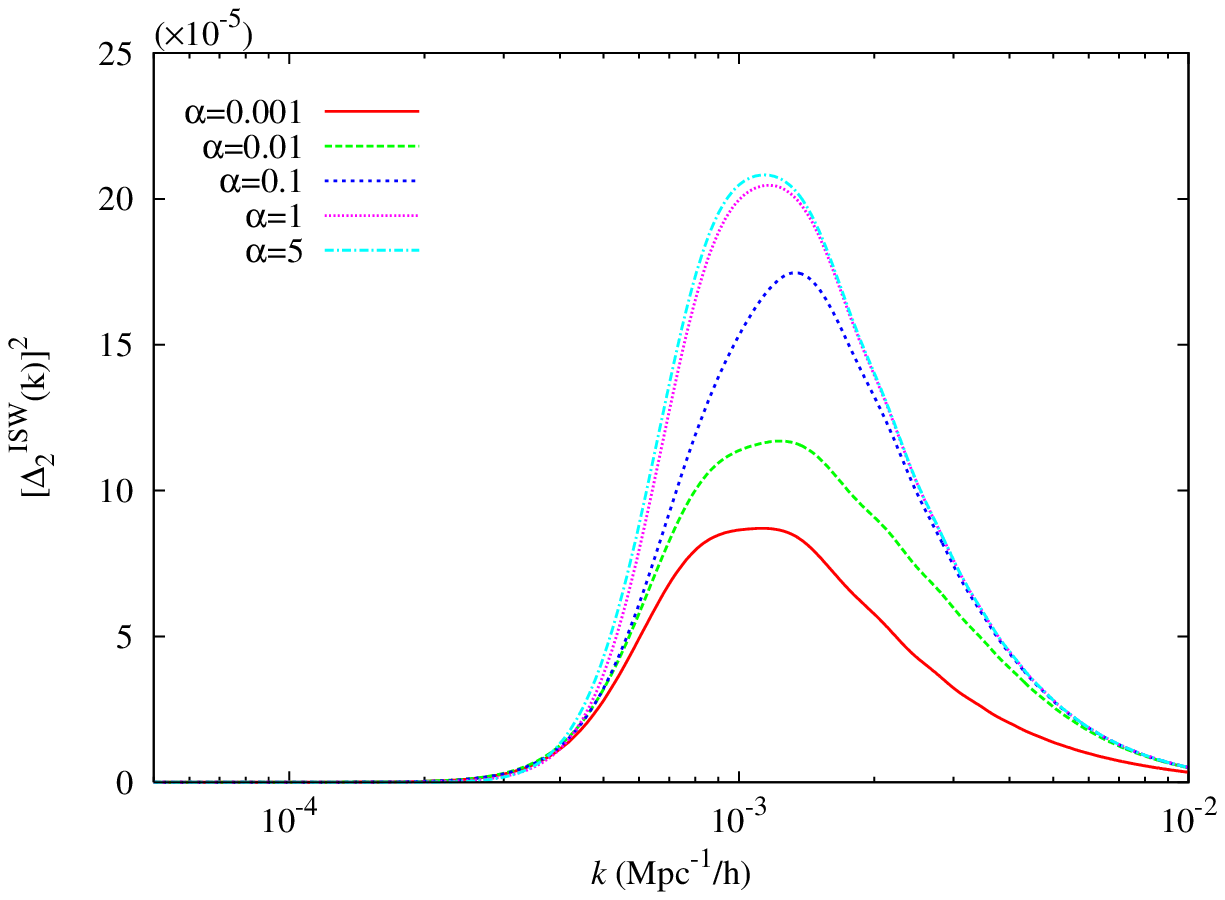}}
\end{minipage}
\begin{minipage}[m]{0.5\linewidth}
\rotatebox{0}{\includegraphics[width=0.9\textwidth]{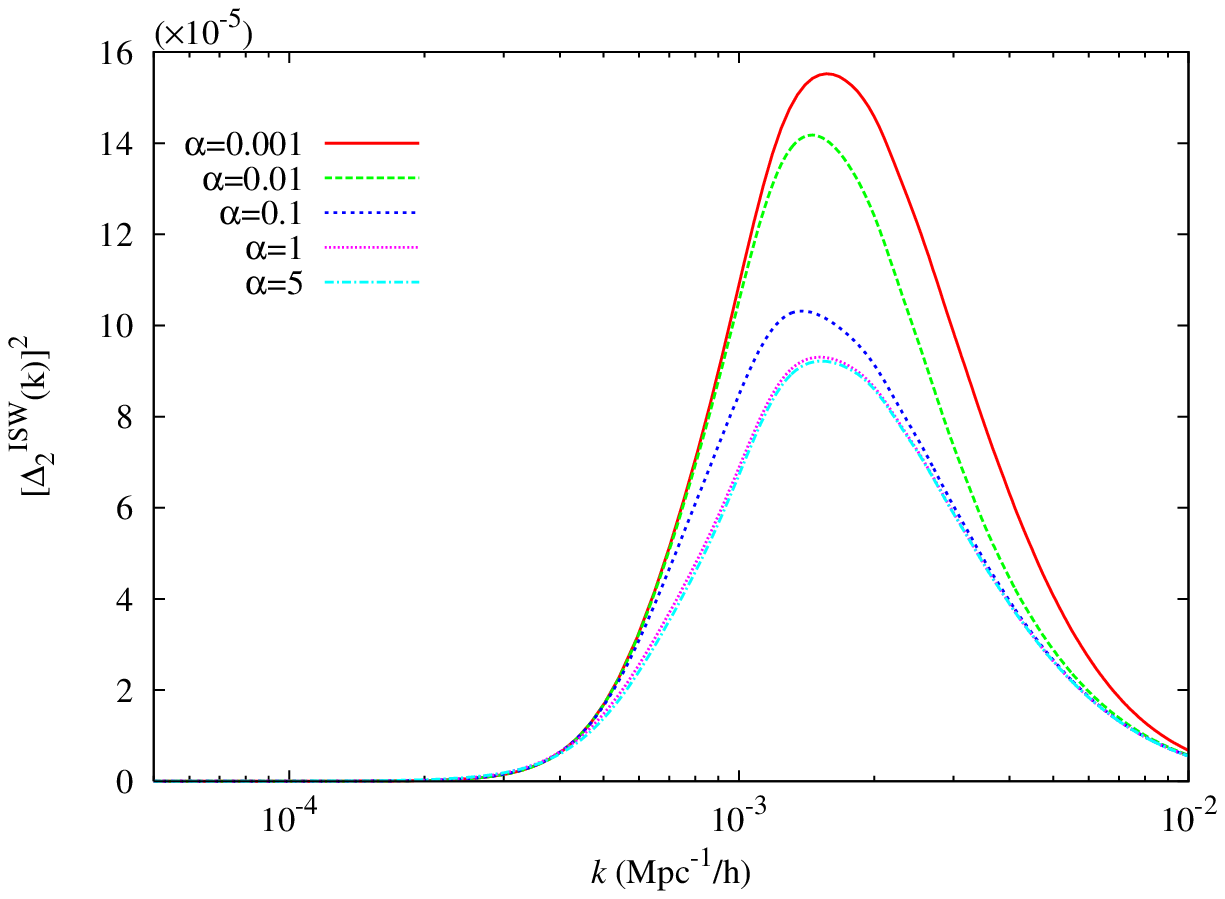}}
\end{minipage}
\caption{Contribution from the ISW effect on the quadrupole
  $\left[ \Delta_2^{\rm ISW} \right]^2$ are shown for several values
  of $\alpha$ for the cases with $w_X=-0.8$ (left) and $-1.2$ (right).
}
\label{fig:7}
\end{figure}

\begin{figure}[htb]
\begin{minipage}[m]{0.5\linewidth}
\rotatebox{0}{\includegraphics[width=0.9\textwidth]{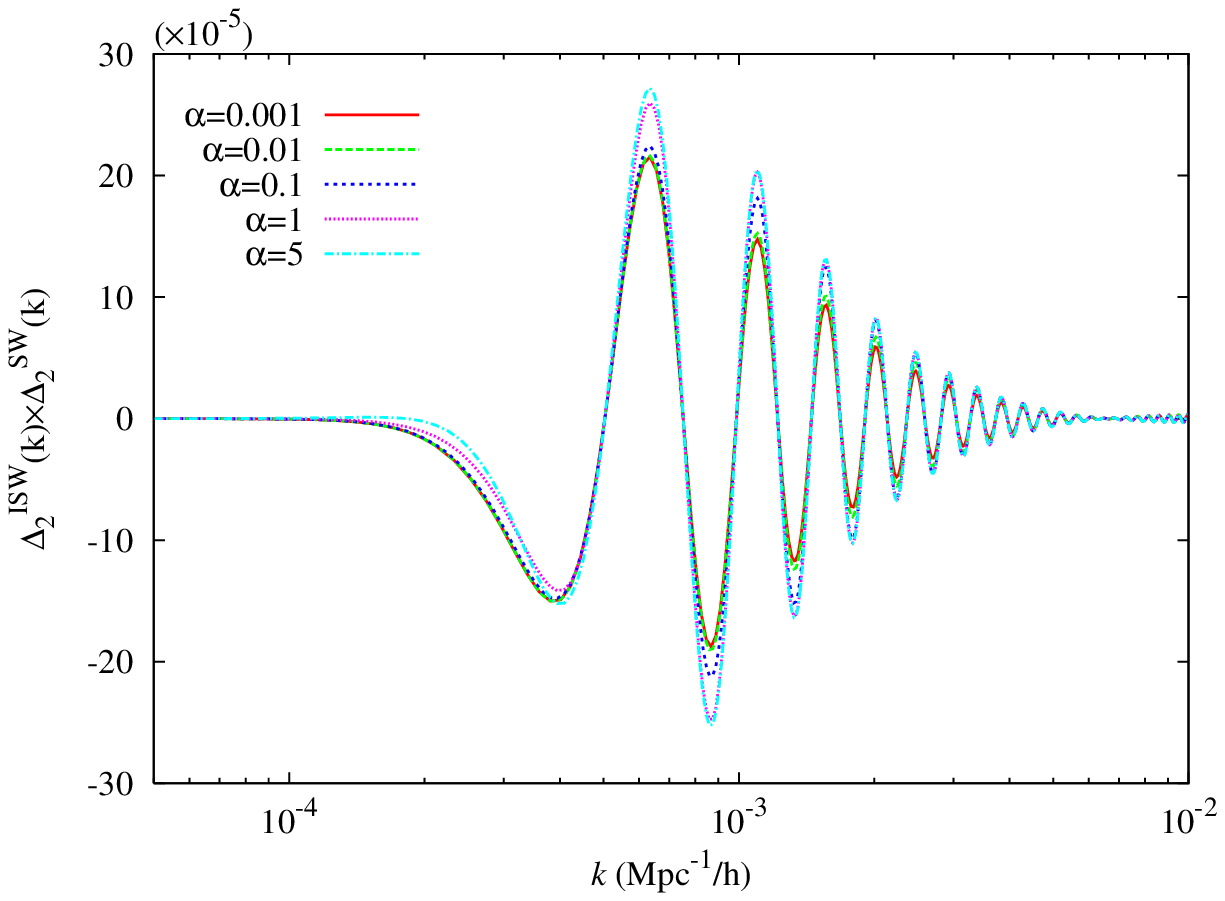}}
\end{minipage}
\begin{minipage}[m]{0.5\linewidth}
\rotatebox{0}{\includegraphics[width=0.9\textwidth]{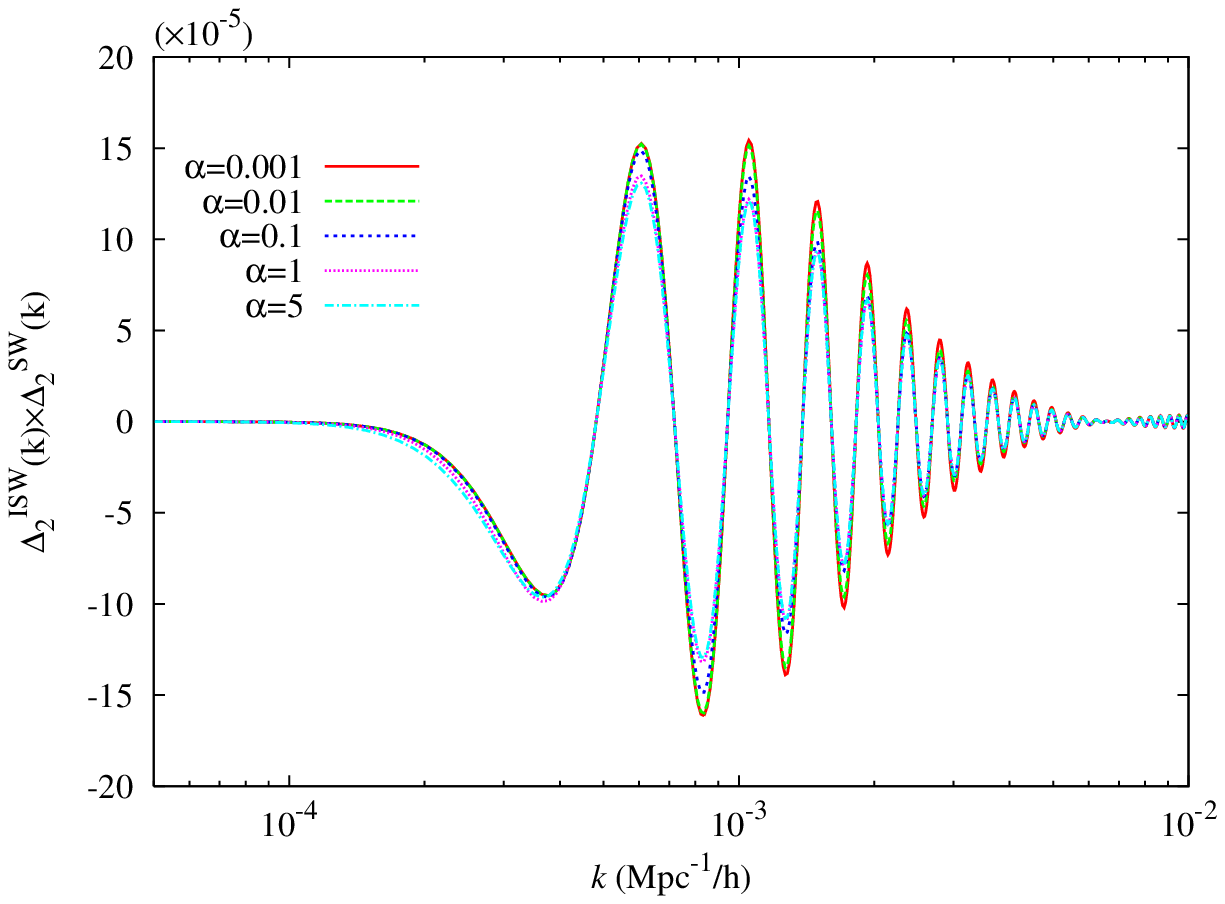}}
\end{minipage}
\caption{Contribution from the ISW effect on the quadrupole $
  \Delta_2^{\rm ISW} \times \Delta_2^{\rm SW}$ are shown for several
  values of $\alpha$ for the cases with $w_X=-0.8$ (left) and $-1.2$
  (right).  }
\label{fig:8}
\end{figure}

Here we should make a comment on a nontrivial cancellation between the
ISW effect and the cross correlation between the ISW and Sachs-Wolfe (SW)
effect.  Although we argued that larger values of $\alpha$ give the
larger ISW effect for the case with $w_X >-1$, 
the lines of $\alpha=1$ and $5$ in the top left
panel of Fig.~\ref{fig:4} do not obey this rule. In fact, this
is because the ISW and SW effects can nontrivially cancel those
effects in some cases. We discuss this issue briefly.  For this
purpose, we make use of the transfer function on large scales.
Formally, $C_l$ is written as
\begin{equation}
C_l = 4\pi \int \frac{dk}{k} P_{\mathcal{R}} (k) | \Delta_l(k)|^2~.
\end{equation}
where $P_{\mathcal{R}}(k)$ is the primordial power spectrum and
$\Delta_l (k)$ represents the transfer function for a
multipole moment $l$ at present time.  On large scales, $\Delta_l(k)$ are mostly
determined by the contribution from the SW and ISW effect, thus it can
be written as
\begin{equation}
\Delta_l(k)  = \Delta_l^{\rm SW}(k) + \Delta_l^{\rm ISW}(k)~.
\end{equation}
In particular, the contribution from the ISW effect is given by
\begin{equation}
\Delta_l^{\rm ISW}(k)  = \int d\eta e^{-\tau} \left( \Phi' + \Psi' \right) j_l [k(\eta - \eta_0)]~,
\end{equation}
where $\tau$ is the optical depth along the line of sight, $\Phi$ is
the gravitational potential and $j_l$ is the spherical Bessel
function.  We plot $|\Delta_{2}^{\rm ISW}|^2$ as a function of wave
number $k$ for several values of $\alpha$ in Fig.~\ref{fig:7} for
the case with $w_X=-0.8$. As seen from the figure, $|\Delta_{2}^{\rm
  ISW}|^2$ becomes larger as $\alpha$ increases.  In fact, as
mentioned above, the quadrupole $C_2$ for $\alpha=1$ is larger than
that for the case with $\alpha=5$ although the ISW effect is larger
for $\alpha=5$. Thus, at first sight, it looks contradictory.
 However, this comes from a non-trivial cancellation
between the contributions from the ISW effect $|\Delta_2^{\rm ISW}|^2$
and the cross correlation between the SW and ISW effects
$\Delta_2^{\rm SW} \times \Delta_2^{\rm ISW}$. In
Fig.~\ref{fig:8}, we also plot $\Delta_2^{\rm SW} \times
\Delta_2^{\rm ISW}$ as the same manner as Fig.~\ref{fig:7}.  When
$\alpha$ is larger, the cross correlation becomes also large in
amplitude as seen from the figure. However, notice that, since the
cross correlation of the SW with the ISW effects can give a negative
contributions to the transfer function.  Thus, for some cases, it
gives smaller quadrupole amplitude as a whole effect. This is the
reason why the low multipoles for the case with $\alpha=1$ is larger
than that for $\alpha=5$ even though the contribution from the ISW
effect monotonically becomes larger as increasing $\alpha$.
For reference, we also show the same plot for the case with $w_X = -1.2$
on the right panel in Figs.~\ref{fig:7} and 
\ref{fig:8}.


\section{Constraints from current cosmological observations}

In this section, we show the constraints from current cosmological
observations on the energy density and equation of state of dark
energy .

To include the effects of non-adiabatic
pressure fluctuation and anisotropic stress of dark energy, we
calculate theoretical 
angular power spectrum by CAMB code modified according to the details in
section 2. Because these effects change the power at larger angular
scales, the amplitude and the shape of the power spectrum at low
multipoles can be used in principle to put constraints on these
fluctuation properties. In fact, however, we cannot find any
significant limits on them because of the large errors due to the cosmic 
variance at low multipoles. Therefore, we conclude that the current
cosmological observations cannot put tight constraints on the $\alpha$
or $c^2_{s}$ parameters. (The forecast of observational constraint on
$c^2_s$ from planning galaxy surveys can be found in
\cite{Takada:2006xs}.)

Even though one cannot give limits on these parameters themselves, it
does not mean that they are irrelevant when considering the constraints
on the nature of dark energy.  
It can happen that one obtains different limits on the cosmological
parameters such as equation of state of dark energy $w_X$ if one takes
different parameters on the fluctuation property of dark energy such as
effective sound speed $c^2_{s}$ and/or anisotropic stress $\alpha$. 
Many observational proposals aim to determine the equation of state
parameter of dark energy $w_X$ as precise as possible, even up to $\sim 1$\%
level \cite{Albrecht:2006um}, because the information about $w_X$ is
important to pin down a model for dark energy, and also to determine the
future of universe \cite{Chiba:2005er}.
It is therefore important to investigate how much  different
assumptions on the fluctuation properties of dark energy can lead the
different observational limits on the equation of state of dark energy.

To study this explicitly, we made likelihood analysis to put
constraint on $w_X$ with several different assumptions on $c^2_s$ and
$\alpha$. The likelihood functions we calculate are based on WMAP three year
data \cite{Hinshaw:2006ia,Page:2006hz} and the 2dF Galaxy Redshift
Survey \cite{Cole:2005sx}.  To make analysis general enough, one
has to vary the other cosmological parameters which also affect the shape of
the CMB and matter power spectra. In order to do so effectively, we
follow Markov chain Monte Carlo (MCMC) approach \cite{Lewis:2002ah}, and
explore the 
likelihood in seven dimensional 
parameter space, namely, $\theta$ (the ratio of the sound horizon to the
angular diameter distance at last scattering), $A_s$ (amplitude of
primordial perturbation), $\Omega_b h^2$, $\Omega_m h^2$, $n_s$,
$\tau$, and $w_X$. Here the Hubble parameter $h$ is 
derived from $\theta$ \cite{Kosowsky:2002zt} and the dark energy density
such that the universe is spatially flat.

\begin{figure}
\begin{minipage}[m]{0.5\linewidth}
\rotatebox{0}{\includegraphics[width=0.9\textwidth]{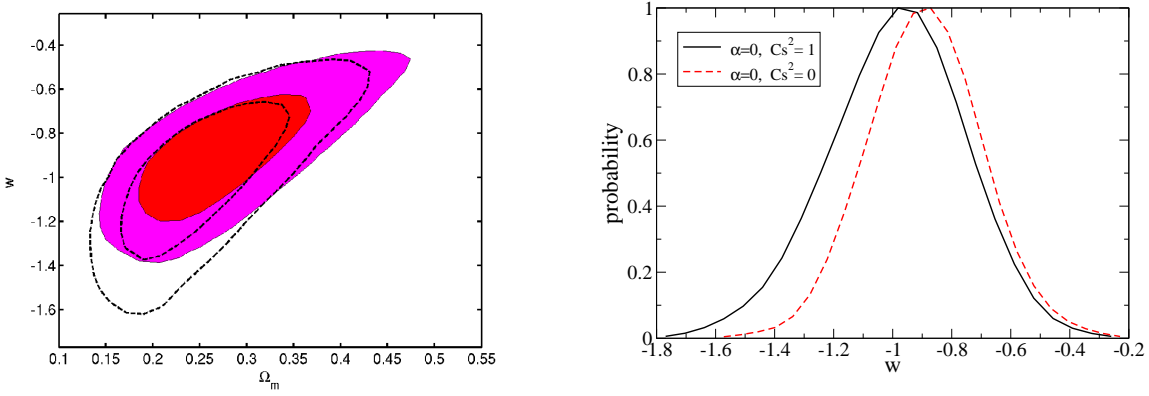}}
\end{minipage}
\begin{minipage}[m]{0.5\linewidth}
\rotatebox{0}{\includegraphics[width=0.9\textwidth]{fig9-2.eps}}
\end{minipage}
\caption{(Left) Allowed regions of 1$\sigma$ (red region) and 2$\sigma$
  (purple region) are shown for the case with $c_s^2 = 0$ and $\alpha
  =0$.  For comparison, allowed regions for the case with $c_s^2=1$
  and $\alpha =0$ are also shown in black dashed lines. (Right) Marginalized probability distribution of the equation of state
 parameter $w_X$ with different assumptions on the fluctuation properties
 of dark energy. Black line corresponds to the case with
 $(c_s^2,\alpha)=(1,0)$ and red dashed line with $(c_s^2,\alpha)=(0,0)$.
}
\label{fig:cs0_alpha0}
\end{figure}

In the left panel of Fig.~\ref{fig:cs0_alpha0}, marginalized 1 and 2$\sigma$ allowed
regions in the $\Omega_m - w_X$ plane from WMAP three year data
are shown for the case with $c_s^2 = 0$ and $\alpha =0$. For comparison,
the case with $c_s^2 = 1$ 
and $\alpha =0$ is also shown in the figure.  As seen from the figure,
the constraint on $w_X$ for the case with $c_s^2=0$ and $\alpha=0$
becomes severer compared to the other case. This is due to the fact
that the ISW effect becomes significant as $w_X$ decreases to 
negative value when $w_X < -1$, which makes the fit to the data worse
at low multipole region. This results indicates that the perturbation
nature of dark energy fluid affects the determination of the equation
of state in some cases.
The marginalized one-dimensional probability distribution of $w_X$ is also 
shown in the right panel of 
Fig.~\ref{fig:cs0_alpha0}. We found that the confidence region of
$w_X$ can differ by $\lesssim 10$\% between the models with
$(\alpha,c_s^2)=(0,0)$ and $(0,1)$. 

\begin{figure}
\begin{minipage}[m]{0.5\linewidth}
\rotatebox{0}{\includegraphics[width=0.9\textwidth]{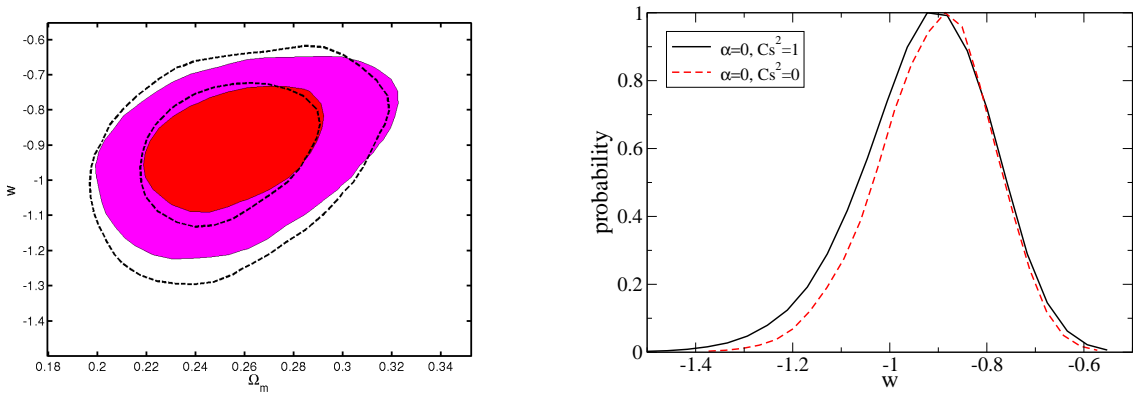}}
\end{minipage}
\begin{minipage}[m]{0.5\linewidth}
\rotatebox{0}{\includegraphics[width=0.9\textwidth]{fig10-2.eps}}
\end{minipage}
\caption{The same as Fig.~\ref{fig:cs0_alpha0} but we also use 
the data from 2dF in addition to WMAP three year data.
}
\label{fig:cs0_alpha0_wPk}
\end{figure}

In the left panel of Fig.~\ref{fig:cs0_alpha0_wPk} 
we show marginalized 1 and 2$\sigma$ allowed regions in the
$\Omega_m - w_X$ plane obtained by combining the information from the
matter power spectra by 2dF galaxy survey with WMAP three year
data. Although the data 
reduce the allowed region in the $\Omega_m - w_X$ plane,
it can still be found that there is a significant difference between the
models with 
different assumptions on the fluctuation properties of dark energy.
Similar analysis and conclusion are done and 
derived by \cite{Pietrobon:2006gh} for a fluid dark energy model
parameterized by $w$ and $c_s^2$, by cross-correlating 
the map from WMAP three year data with that from NRAO VLA sky survey
data (see also \cite{Hu:2004yd}).
The present work differs from them in that we analyzed different
observational data set, included viscosity parameter $\alpha$, and
relaxed the other cosmological parameters to be varied while they fixed
cosmological parameters except for the dark energy parameters.

\begin{figure}[htb]
\begin{minipage}[m]{0.5\linewidth}
\rotatebox{0}{\includegraphics[width=0.9\textwidth]{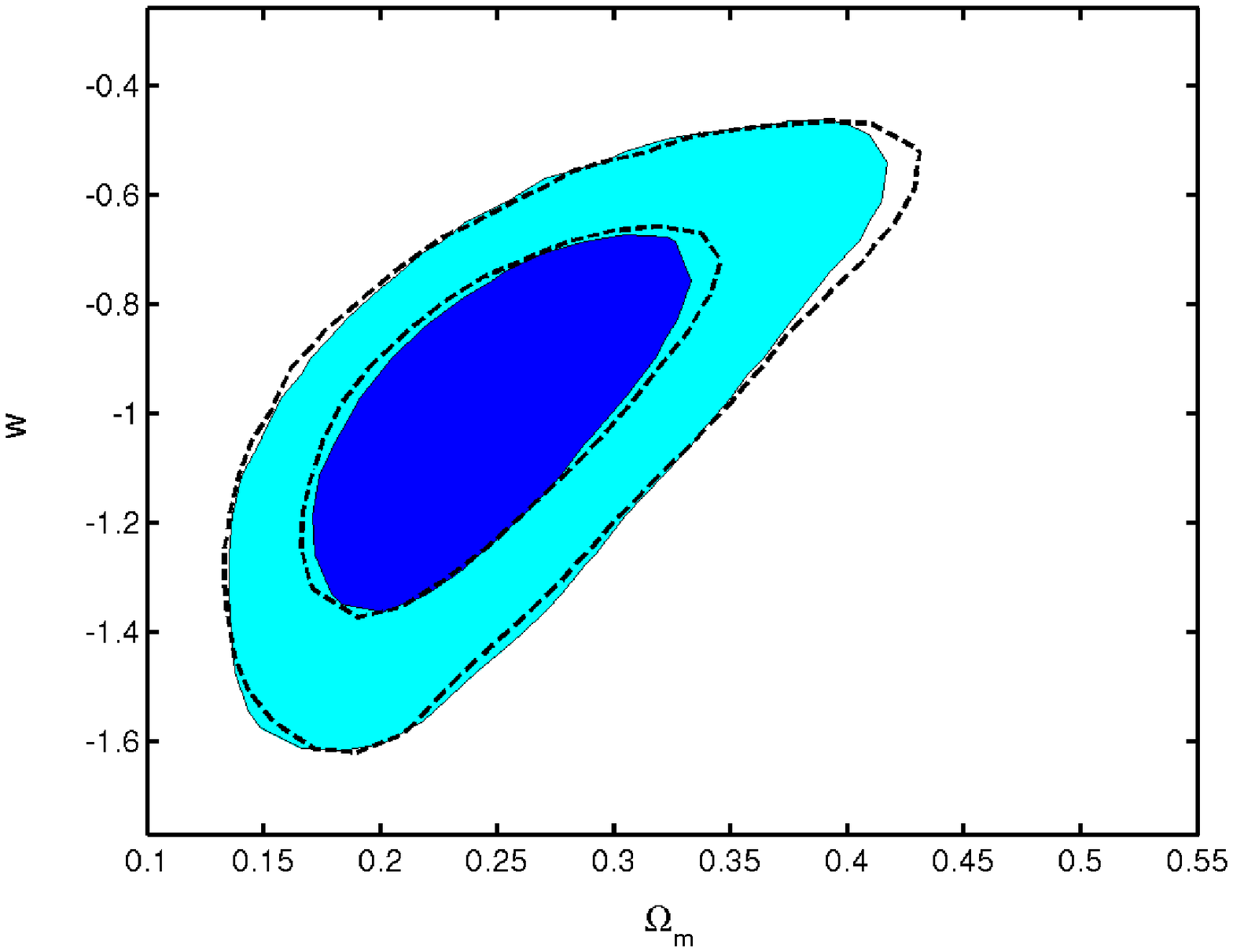}}
\end{minipage}
\begin{minipage}[m]{0.5\linewidth}
\rotatebox{0}{\includegraphics[width=0.9\textwidth]{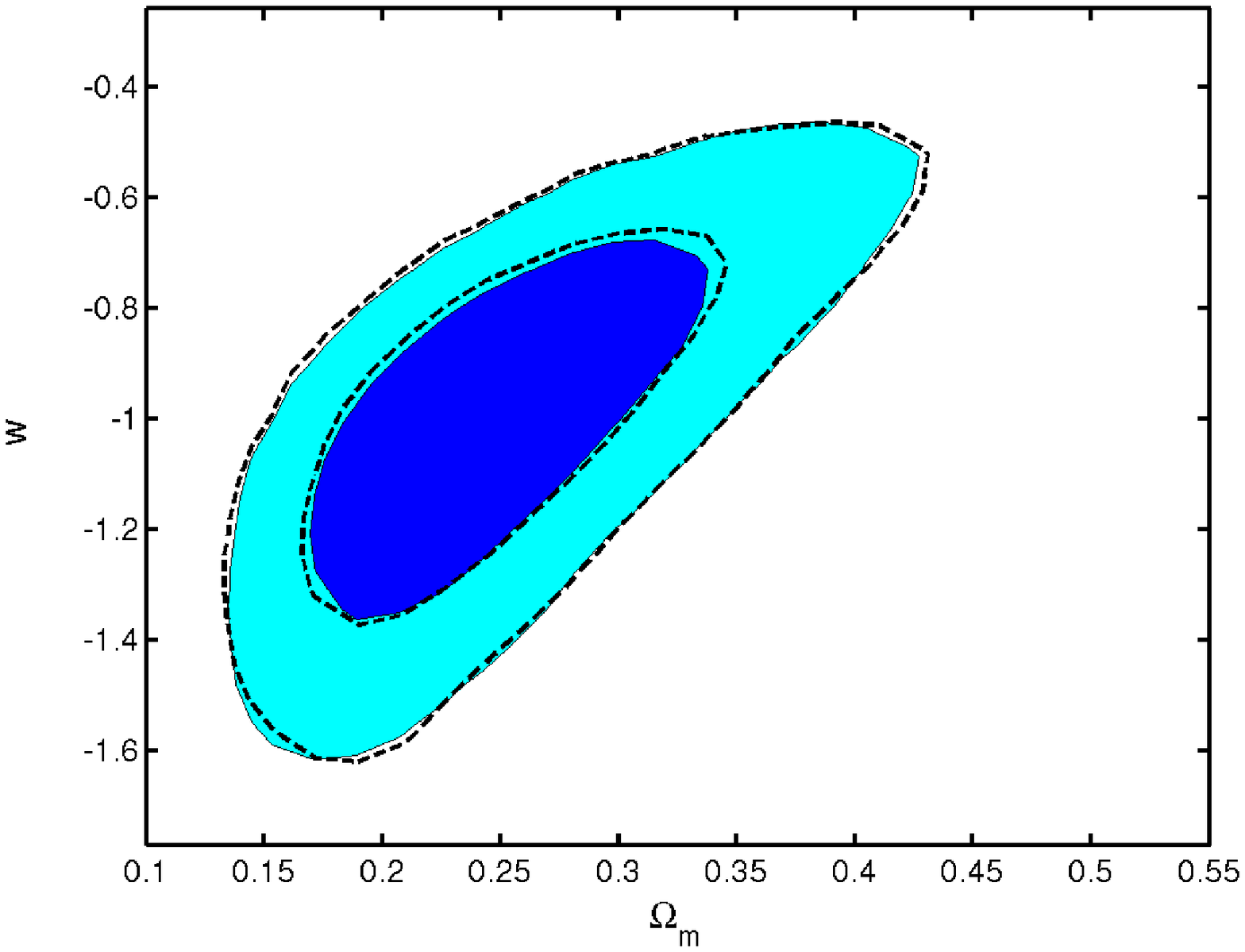}}
\end{minipage}
\caption{Allowed regions of 1$\sigma$ (blue region) and 2$\sigma$
  (cyan region) are shown for the case with $c_s^2= 0$ and $\alpha =0.6$
(left) and $c_s^2= 1$ and $\alpha =0.6$ (right). For comparison, allowed regions for the case with $c_s^2=1$
  and $\alpha =0$ are also shown in black dashed lines. }  
\label{fig:other_cs_alpha}
\end{figure}

\begin{figure}[htb]
\begin{minipage}[m]{0.5\linewidth}
\rotatebox{0}{\includegraphics[width=0.9\textwidth]{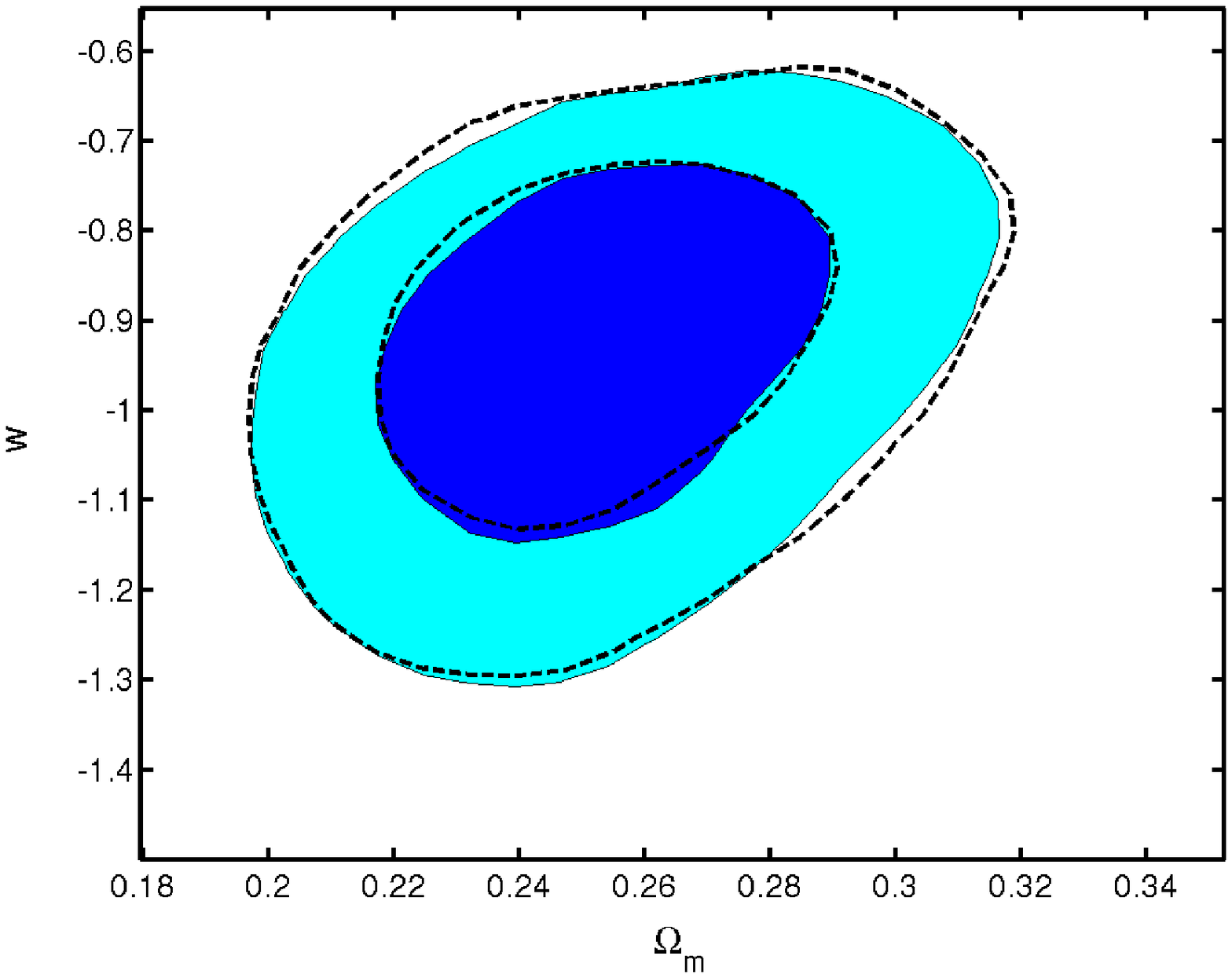}}
\end{minipage}
\begin{minipage}[m]{0.5\linewidth}
\rotatebox{0}{\includegraphics[width=0.9\textwidth]{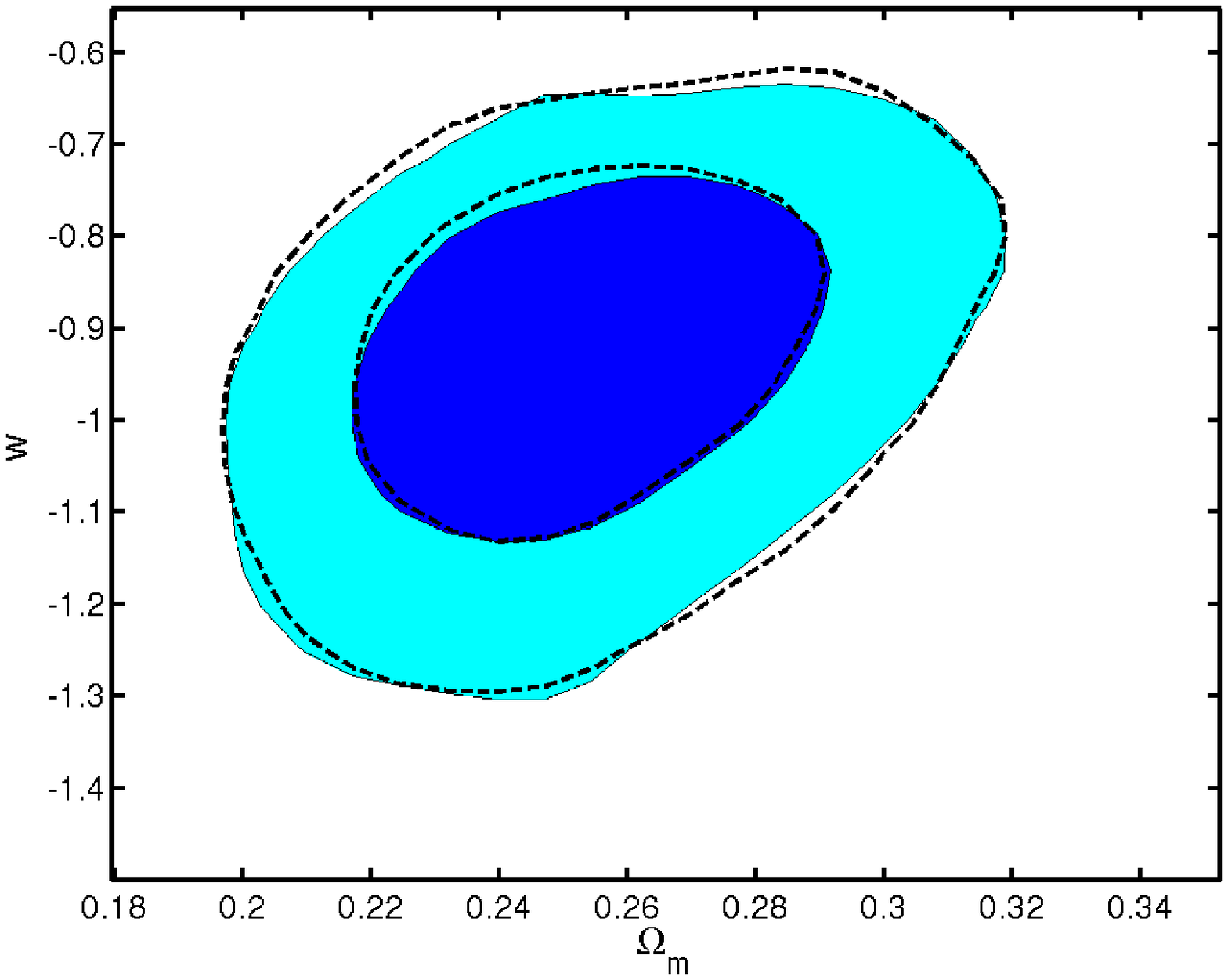}}
\end{minipage}
\caption{The same as Fig.~\ref{fig:other_cs_alpha} but we also use 
the data from 2dF in addition to WMAP three year data.}  
\label{fig:other_cs_alpha_wPk}
\end{figure}

In Fig.~\ref{fig:other_cs_alpha}, marginalized 1 and 2$\sigma$ allowed
regions are shown for the case with $(c_s^2, \alpha) = (0,0.6)$ (left) and
$(1,0.6)$ (right).  In fact, large value of $c_s^2 / \alpha$ makes any
dependence of $w_X$ at low multipoles insignificant. Furthermore, we
found that when $\alpha \sim 1$  the result is 
insensitive to the value of $c_s^2$ because of the fact that the
fluctuation amplitude at low multipoles is independent from $c_s^2$ in
such cases (see Fig. \ref{fig:6}). 
Thus in this case,
the constraint on $\Omega_m - w_X$ plane is almost unchanged.
We also show the same plot for the case where we use the data from 2dF 
in addition to WMAP three year data in 
Fig.~\ref{fig:other_cs_alpha_wPk}. 
As seen from the figure and easily expected, 
the constraints become severer but the differences 
among different assumptions on $c_s^2$ and $\alpha$ are small.

\section{Conclusions and discussions}

We discussed the effects of a generalized dark energy on cosmic
density fluctuations and studied cosmological constraints on dark
energy in that setting.  When one considers a general type of dark
energy, one can also include a possible anisotropic stress of dark
energy, which has not been investigated much in the literature.  We
studied the effects of the anisotropic stress characterized by the
viscosity parameter $\alpha$ along with the non-adiabatic pressure
fluctuation
usually described by the speed of sound $c_s^2$. First we discussed
the effects of them on the gravitational potential and argued how the
values of $\alpha$ and $c_s^2$ affect it.  As shown, in section
\ref{sec:effect}, the anisotropic stress of dark energy can affect the
ISW effect in a similar way with the non-adiabatic pressure
fluctuation.  We also 
explicitly showed that there is some degeneracy between the effects of
$\alpha$ and $c_s^2$ on the power at low multipoles of CMB angular power
spectrum.

We also explore constraints on dark energy from recent cosmological
observations assuming a generalized dark energy.  As shown in
Fig.~\ref{fig:cs0_alpha0}, the constraint on the equation of state can
be changed depending on the parameters $c_s^2$ and $\alpha$ which
describes the perturbation nature of dark energy.  This is because
fluctuation on large scales are affected by the nature of dark energy
fluctuation, i.e., depending on $\alpha$ and $c_s^2$, through the ISW
effect.  This analysis may indicate that cosmological constraint on
dark energy such as the equation of state should be carefully done by
taking the perturbation nature of dark energy into account.

\bigskip
\bigskip

\noindent
{\bf Acknowledgments:}  KI acknowledges the support by a Grant-in-Aid
for the Japan Society for the Promotion of Science.

\end{document}